\documentclass[aps, prd, twocolumn, nofootinbib, floatfix]{revtex4}
\usepackage{subfigure}
\usepackage{graphicx}
\usepackage{dcolumn}
\usepackage{epsfig}
\usepackage{amsmath}
\usepackage{amsfonts}
\usepackage{amssymb}
\usepackage{color}
\usepackage{hyperref}
\usepackage{textcomp}
\newcommand{\gsim}{\;\mbox{\raisebox{-0.5ex}{$\stackrel{>}{\scriptstyle{\sim}}$}}\;}

\newcommand{\be}{\begin{equation}}
\newcommand{\ee}{\end{equation}}
\newcommand{\bea}{\begin{eqnarray}}
\newcommand{\eea}{\end{eqnarray}}
\newcommand{\nn}{\nonumber}

\begin{document}

\title{Laguerre reconstruction of the correlation function on Baryon Acoustic Oscillation scales 
}

\author{Farnik Nikakhtar${}^1$}
\email{farnik@sas.upenn.edu}
\author{Ravi K.~Sheth${}^{1,2}$}
\author{Idit Zehavi${}^3$}

\affiliation{${}^1$Department of Physics and Astronomy, University of Pennsylvania, 209 S. 33rd St., Philadelphia, PA 19104 -- USA}
\affiliation{${}^2$Center for Particle Cosmology, University of Pennsylvania, 209 S. 33rd St., Philadelphia, PA 19104 -- USA}
\affiliation{${}^3$Department of Physics, Case Western Reserve University, Cleveland, OH 44106-7079 -- USA}

\date{\today}

\begin{abstract}
  The baryon acoustic oscillation feature can be used as a standard cosmological ruler.  In practice, for sub-percent level accuracy on the distance scale, it must be standardized.  The physical reason why is understood, so we use this to develop an algorithm which improves the estimated scale.  The algorithm exploits the fact that, over the range of scales where the initial correlation function is well-fit by a polynomial, the leading order effects which distort the length of the ruler can be accounted for analytically.  Tests of the method in numerical simulations show that it provides simple and fast reconstruction of the full shape of the BAO feature, as well as subpercent determination of the {\em linear point} in the correlation function of biased tracers with minimal assumptions about the underlying cosmological model or the nature of the observed tracers. Our results also suggest that, for least squares estimators of the correlation function, half-integer generalized Laguerre functions are a particularly useful choice.
\end{abstract}


\maketitle

\newcommand{\ste}[1]{\textcolor{red}{\textbf{\small[Ste: #1]}}}


\section{Introduction}\label{intro}
Baryon acoustic oscillations from the early universe imprint a characteristic feature in the spatial distribution of matter even at much later times \cite{peeblesYu,esw2007}.  This feature -- a peak and dip in the two-point correlation function on scales of order 150~Mpc (comoving) -- has been used to constrain the background cosmological model via the distance-redshift relation \cite{sanchezDR12}, and there is hope that it can also be used to constrain the growth of clustering \cite{alamDR12}.  

However, on BAO scales, the evolved two-point correlation function, even of unbiased tracers, differs in shape from the unbiased linear correlation function \cite{rpt, bkPeaks}.  The difference is particularly dramatic near the peak and dip of the BAO feature, and has motivated a number of algorithms for `reconstructing' the shape of the BAO feature \cite{recPW, recIterate, recHE}.  Most of these involve modifying the positions of the tracer particles -- e.g. dark matter halos in simulations or galaxies in observations -- so as to return them to their `linear theory' values. These `density field reconstruction' approaches are effective, but are computationally expensive and closely tied to an assumed fiducial cosmological model.  More recent algorithms, e.g., the extended fast action minimisation method \cite{eFAM2019}, and the fast semi-discrete optimal transport algorithm \cite{royaMAK}, are more computationally efficient. In what follows, we outline a rather different approach which is much cheaper and less tied to a cosmological model.  We use the Linear Point (LP) -- the scale that lies midway between the peak and dip, which previous work has shown can be used as a standard cosmological ruler \cite{PaperI, PRDmocks, PRLboss, LPnus, LPruler} -- to quantify the accuracy and precision of our reconstruction algorithm.

Section~\ref{sec:method} describes our method.
Section~\ref{sec:results} shows our results.
Section~\ref{sec:stats} discusses how they can be used to set constraints on the distance scale.  
Section~\ref{sec:disc} summarizes.  Additional technical details are provided in three Appendices.  Some of these details illustrate the power of using a polynomial basis for describing the shape of $\xi$, a point recently made by \cite{smallr, Krolewski21} regarding the small-scale regime which is not the focus of our study.  

\section{Methodology}\label{sec:method} 
We describe our methodology in three steps.  The first two treat the simplest case, which may be all that is necessary for dark matter:  following \cite{rpt}, these are sometimes called the `convolution' and `mode-coupling' terms.  We use these to set up notation and outline the underlying philosophy of the approach.  The third adds complications that may be necessary for treating biased tracers.  What results is a three step algorithm which begins with fitting any observed correlation function to Eq.(\ref{xiNLb}).  

In section \ref{sec:results}, we use numerical simulations to validate our methodology.  Hence, all the figures in this section are for the same background cosmological model as the simulations.  


\subsection{Evolved \texorpdfstring{$\xi_{\mathrm{NL}}$}{} as convolution of \texorpdfstring{$\xi_{\mathrm{L}}$}{}}
Our starting point is motivated by \cite{Bharadwaj1996} and \cite{rpt}, and states that the evolved pair correlation function is related to that predicted by linear theory (i.e. the initial one multiplied by a growth factor) by a convolution:  
\begin{equation}
 \xi_{\rm NL}(\mathbf{s}) \approx \int d\mathbf{r}\,\xi_{\rm L}(\mathbf{r})\,G(\mathbf{s-r}|\Sigma).
 \label{xiNLconv}
\end{equation}
The approximate sign here is because we are ignoring what are sometimes called `mode coupling' terms that are known to be small \cite{rpt,bkPeaks}.  We discuss how to include them later. We have used $G$ to indicate that the smearing kernel is Gaussian; $\Sigma$ is its rms (in Mpc).  While its exact value is not important for the argument which follows, it is useful to know that 
 $\Sigma^2\approx\int dk P_{\rm L}(k)/3\pi^2$,
where $P_{\rm L}(k)$ is the linear theory power spectrum \cite{rpt}.  For cosmological models of current interest, $\Sigma$ is proportional to the linear theory growth factor $D(z)$ and is substantially smaller than the BAO scale.

\begin{figure}[t]
 \centering
 \includegraphics[width=0.9\hsize]{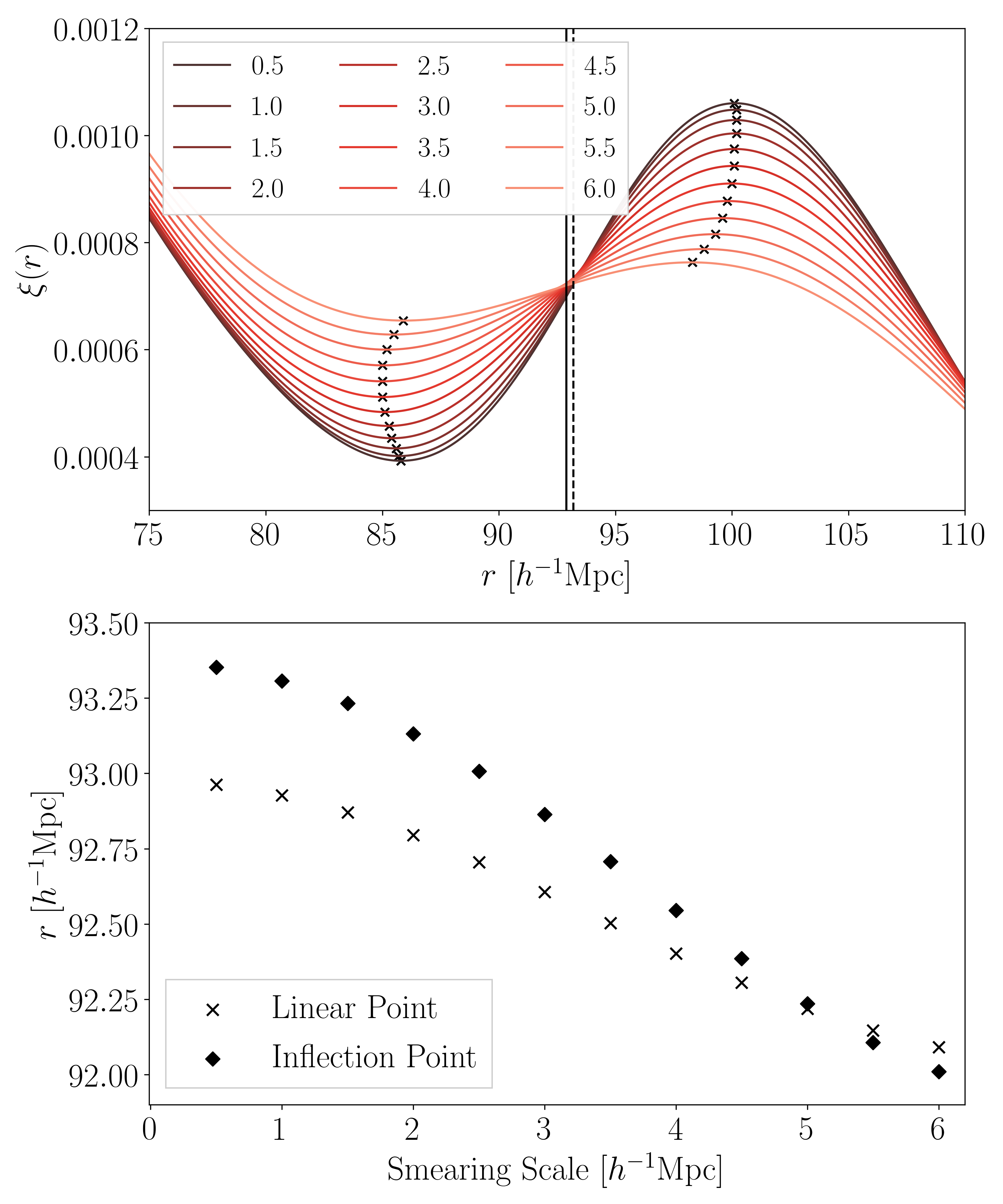}
 \caption{\label{fig:smoothing} 
  Effect of smearing on the shape of the correlation function. In top panel, vertical black solid and dashed lines show $r_{\rm LP}$ and $r_{\rm infl}$ in the unsmoothed $\xi_{\rm L}$. Curves show how the shape of $\xi$ changes as the smoothing increases (Eq.~\ref{xiNLconv}), and crosses show the peak and dip positions for each smeared correlation function. Bottom panel shows $r_{\rm LP}$ and $r_{\rm infl}$ in the smeared correlation function:  $r_{\rm LP}$ is slightly more robust to smearing.}
\end{figure}

The top panel of Fig.~\ref{fig:smoothing} shows the effect of smoothing on the shape of the correlation function.  The most obvious effect is that smoothing smears out the peak and dip.  Crosses show the peak and dip positions for each smeared correlation function:  they change with smearing scale, but it is apparent that their average may be more stable. Indeed, as first noticed by \cite{PaperI}, the linear point scale
\begin{equation}
 r_{\rm LP} \equiv \frac{r_{\rm peak} + r_{\rm dip}}{2}
 \label{eq:rLP}
\end{equation}
is almost unaffected by the smearing.  
We will also discuss the inflection point $r_{\rm infl}$ which is the scale between the peak and dip where $d^2\xi/dr^2 = 0$.  The two scales are very close:  The vertical black solid and dashed lines show $r_{\rm LP}=93h^{-1}$Mpc and $r_{\rm infl}=93.4h^{-1}$Mpc for the initial unsmoothed $\xi_{\rm L}$.
The stability of $r_{\rm infl}$ to evolution is easier to understand, but $r_{\rm LP}$ turns out to be slightly more stable \cite{PaperI}.

The smearing is expected to increase with time
\cite[e.g.][]{rpt}.
The bottom panel shows the linear and inflection points as a function of smearing scale.  For dark matter at $z=0.5$ we expect $\Sigma = 4.6h^{-1}$Mpc, for which Fig.~\ref{fig:smoothing} indicates the measured linear point will be changed to $92.25h^{-1}$Mpc from the unsmoothed $93h^{-1}$Mpc scale.  While this $0.75h^{-1}$Mpc change/shift is much smaller than the amount by which the peak and dip positions themselves change, it is comparable to the precision with which the next generation of sky surveys will measure this scale.  This is why \cite{PaperI} recommended that a 0.5 percent correction be applied to any measured value (i.e. multiply the measured value by 1.005).  Since the shift may depend on tracer particle type -- and we show below that it does -- we will {\em not} do this.  Rather, our goal is to recover the linear theory (i.e. unsmoothed) values of $r_{\rm LP}$ and $r_{\rm infl}$ from measurements of the evolved correlation function, assuming Eq. (\ref{xiNLconv}) is accurate.  Fig. 4 of \cite{LPnus} shows that Eq. (\ref{xiNLconv}) indeed provides a good description of the evolution of the peak and dip scales in simulations.

\subsection{Analytic (de)convolution}
Since $\xi_{\rm L}$ is isotropic,
Eq.~(\ref{xiNLconv}) becomes
\begin{equation}
  \xi_{\rm NL}(s)= \int_0^\infty \frac{dr\,r^2}{\Sigma^3}\,\frac{{\rm e}^{-(r^2+s^2)/(2\Sigma^2)}}{\sqrt{2\pi}}\,2\frac{\sinh(rs/\Sigma^2)}{rs/\Sigma^2}\,\xi_{\rm L}(r).
  \label{xiNLiso}
\end{equation}
The terms other than $\xi_{\rm L}$ in the integral define a noncentral-Chi distribution in $r/\Sigma$ with 3 degrees of freedom, with noncentrality parameter $s/\Sigma$, so it is useful to write Eq.(\ref{xiNLiso}) as 
\begin{equation}
  \xi_{\rm NL}(s) = \int_0^\infty \frac{dr}{\Sigma}\,\chi_3\left(\frac{r}{\Sigma}\Big|\frac{s}{\Sigma}\right)\,\xi_{\rm L}(r).
  \label{xiNLchi3}
\end{equation}

Next, suppose that $\xi_{\rm L}$ can be well approximated by  
\begin{equation}
 \xi_{\rm L}(r) = \sum_{k=0}^n a_k\,(r/\sigma)^k,
 \label{xiLpoly}
\end{equation}
where $\sigma$ is set equal to a fiducial value, as this makes all the $a_k$ dimensionless.
When inserted in Eq.(\ref{xiNLchi3}) this polynomial representation yields $\xi_{\rm NL}$ as a sum over moments of the $\chi_3$ distribution.  If we define $x\equiv s/\Sigma$ then 
\begin{equation}
  \xi_{\rm NL}(s) = \sum_{k=0}^n c_k\,\mu_k(x),\quad {\rm where} \quad
  c_k \equiv a_k\, \left(\frac{\Sigma}{\sigma}\right)^k
  \label{xiNLpoly}
\end{equation}
and
\begin{align}
 \mu_{2n} &= 2n!!\, L_{n}^{(1/2)}(-x^2/2) \nn \\
 \mu_{2n-1} &= (2n-1)!!\,\sqrt{\frac{\pi}{2}}\,L_{n-1/2}^{(1/2)}(-x^2/2).
 \label{chi3-moments}
\end{align}
The $L_{\beta}^{(\alpha)}(z)$ are generalized Laguerre functions, which we discuss more in Appendix~\ref{sec:Lab}.  For integer $\beta$ they are simple polynomials, but otherwise they are complicated functions.  I.e., if $\xi_{\rm L}$ is a polynomial of order $n$, then $\xi_{\rm NL}$ will not be a simple polynomial.  That said, Appendix~\ref{sec:explicit} shows that $\xi_{\rm NL}$ reduces to a simple polynomial in the limit in which the scales of interest are much larger than $\sigma$.  This explains why \cite{PRDmocks} found that a simple polynomial can provide a good fit to $\xi_{\rm NL}$.

The results above suggest that we should:
\begin{enumerate}
 \item Fit Eq.(\ref{xiNLpoly}) --- rather than a simple polynomial --- to the measured $\xi_{\rm NL}$;
 \item Then use the fitted $c_k$ to estimate $a_k = c_k (\sigma/\Sigma)^k$;
 \item Finally, insert these $a_k$ into Eq.(\ref{xiLpoly}) to obtain the `deconvolved' or `reconstructed' shape, which we will sometimes refer to as $\xi_{\rm Lag}$ (for `Laguerre reconstructed $\xi$').
\end{enumerate}
We discuss a few technical details associated with Step 1 in Appendices.  Centering the functions to be fit around a fiducial scale, so as to avoid numerical inaccuracy, is the subject of Appendix~\ref{sec:centered}. How we determine the order of the polynomial and the range of scales over which to fit is the subject of Appendix~\ref{fitDetails}.

Step 2 makes obvious that the reconstruction depends on what one chooses for $\Sigma$ (recall $\sigma$ is just a fiducial value).  So, one way to proceed is to fit $\xi_{\rm NL}$ to Eq.(\ref{xiNLpoly}) assuming $\Sigma$ equals the fiducial value.  At a later stage, one can weight each `reconstruction' by a prior on the fiducial value. We discuss an alternative approach to determining $\Sigma$ in Section~\ref{sec:ba}.

Finally, although we have concentrated on reconstructing the shape of $\xi_{\rm L}$ from the measured $\xi_{\rm NL}$, for LP purposes, one is most interested in the scale which is midway between the peak and dip in $\xi_{\rm Lag}$, or the inflection point between them (i.e., where $\xi_{\rm Lag}''=0$).  Since $\xi_{\rm NL}' = \xi_{\rm Lag}' + (\xi_{\rm NL} - \xi_{\rm Lag})'$ and similarly for $\xi_{\rm NL}''$, the zeros of $\xi_{\rm Lag}'$ are where 
\begin{equation}
  \label{rLPrecon}
  \frac{\partial\xi_{\rm NL}(s)}{\partial\ln s} = \sum_{k=0}^n a_k\,\left(\frac{\partial\mu_k(x)}{\partial\ln x} - kx^k\right)
\end{equation}
rather than where $\xi_{\rm NL}'=0$. The zeros of the above equation give the $s$ which are the peak and dip scales, from which $r_{\rm LP}$ can be obtained (Eq.~\ref{eq:rLP}).


\subsection{Illustration and formal uncertainties}
\begin{figure}
 \centering
 \includegraphics[width=0.95\hsize]{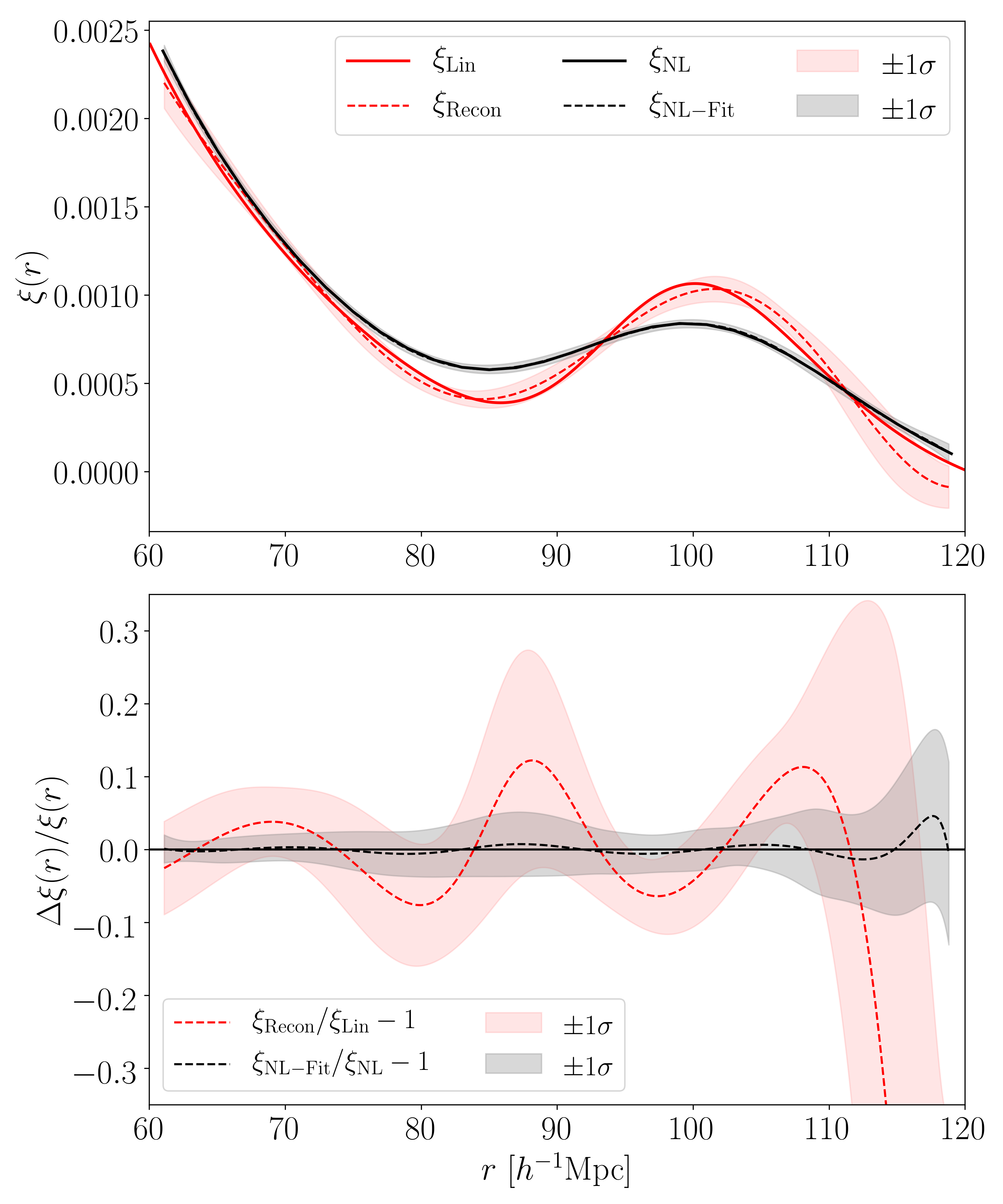}
 \caption{\label{fig:allTheory} 
   Illustration of how Laguerre deconvolution `reconstructs' the shape of the linear theory correlation function $\xi_{\rm L}$.  Top panel:  Red solid curve shows $\xi_{\rm L}$ and black solid curve shows the result of convolving it with a Gaussian kernel of width $4.6h^{-1}$Mpc (i.e. $\xi_{\rm NL}$ of Eq.\ref{xiNLconv}).  Black dashed curve shows the 9th-order Laguerre function which best-fits $\xi_{\rm NL}$, and grey region surrounding it shows the associated uncertainty band (see text).  Dashed red curve and pink region shows the deconvolved correlation function $\xi_{\rm Lag}$ and its associated error band when $\Sigma=\sigma=4.6h^{-1}$Mpc.  The linear theory shape is quite well reconstructed.  Bottom panel:   Fractional differences between the Laguerre fit and $\xi_{\rm NL}$ (black) and the reconstructed $\xi_{\rm Lag}$ and true $\xi_{\rm Lin}$ (red).  Small differences between the Laguerre-fit and $\xi_{\rm NL}$, and the associated uncertainty bands, are amplified by the deconvolution.} 
\end{figure}

Figure~\ref{fig:allTheory} illustrates the method.  In the top panel, the solid red curve shows $\xi_{\rm L}$, and the black solid curve shows $\xi_{\rm NL}$ of Eq.(\ref{xiNLconv}) with $\Sigma=4.6h^{-1}$Mpc.  A black dashed curve, which is barely distinguishable in the top panel, shows the result of fitting a 9th-order Laguerre function to $\xi_{\rm NL}$ over the range $60-120h^{-1}$Mpc.  The fitting takes as input the values of $\xi_{\rm NL}$ in equally-spaced, adjacent but non-overlapping bins of width $3h^{-1}$Mpc, and the error covariance matrix associated with a source density of $6.9\times 10^{-3}$ (Mpc$^{-1} h$)$^3$ in a survey volume of $\sim 50(h^{-1}$Gpc$)^3$.
We estimate the covariance matrix using Eq.(2.8) of \cite{LPnus}, which is taken from \cite{eppur}.

In addition to returning the values of the ten fitted parameters $c_k$, the fitting routine outputs an estimate of the covariance between the fitted $c_k$. It is standard practice to use this to derive uncertainty bounds on the best-fit shape, which we show as a grey band.  The black dashed curve and grey band in the bottom panel show that the fit is quite good. 

The dashed red curve in the top panel shows the result of setting the fiducial smearing value $\sigma$ equal to the actual smearing value $\Sigma$, hence setting $a_k=c_k$ (c.f. Eq.\ref{xiNLpoly}) in Eq.(\ref{xiLpoly}).  The covariance between the fitted $c_k$ results in the one sigma pink band around the red dashed curve.  Clearly, the reconstructed shape is much closer to $\xi_{\rm L}$ than was $\xi_{\rm NL}$.

The red dashed curve and associated pink band in the bottom panel show the fractional difference between this deconvolved or reconstructed shape and the original linear theory curve.  Comparison with the black dashed curve in the bottom panel shows that deconvolution amplifies small inaccuracies in the fit to $\xi_{\rm NL}$. This is consistent with conventional wisdom:  whereas convolution smears out fine-scale details in the original signal, in the process of sharpening them again, deconvolution may also amplify features which are due to noise.  E.g., in the middle of the fitted range, the red dashed curve is like an amplified version of the black dashed curve, but this correspondence is not as tight near -- i.e. within about $\Sigma$ -- the boundaries of the fitted region (again, this is as expected for deconvolution).

For linear point analyses, we are not as interested in the full shape as we are in $r_{\rm LP}$ and $r_{\rm infl}$.  In particular, we would like to know if deconvolution reduces the biases in the inferred scales (c.f. the values associated with $4.6h^{-1}$Mpc in the bottom panel of Fig.~\ref{fig:smoothing}).  If it does, we would like to know if it increases the uncertainties on the reconstructed values.  Following, e.g., \cite{LPnus}, the uncertainty on $r_{\rm LP}$ from $\xi_{\rm NL}$ is the square root of 
\begin{equation}
  \sigma_{\rm LP}^2 = \sum_{i,j} \frac{\partial r_{\rm LP}}{\partial c_i}
       \Big\langle(c_i - \langle c_i\rangle)(c_j - \langle c_j\rangle)\Big\rangle
                             \frac{\partial r_{\rm LP}}{\partial c_j}
\end{equation}
where $r_{\rm LP}$ is that nonlinear combination of the $c_k$ and $\mu_k(x)$ functions which comes from requiring $\xi_{\rm NL}'=0$.  The uncertainty on $r_{\rm LP-recon}$ is given by a similar expression, except that now we have $a_k$ coefficients and the nonlinear combination is from solving Eq.(\ref{rLPrecon}). The analysis for $r_{\rm infl}$ is similar.  

Prior to deconvolving, we find that
 $r_{\rm LP-pre} = 92.19\pm 0.15\, h^{-1}$Mpc; 
increasing this value by a factor of $1.005$ (as \cite{PaperI} advocate) would bring it to within about $0.35h^{-1}$Mpc of the linear theory value of $93h^{-1}$Mpc.  After deconvolving, we find 
 $r_{\rm LP-rec} = 93.01\pm 0.14 h^{-1}$Mpc;
no additional shift is necessary.  Results for $r_{\rm infl}$ are similarly encouraging.  This motivates extending the approach to include additional complications that may arise when working with biased tracers.  

\subsection{Mode-coupling:  Dark matter}
For dark matter, Eq.(\ref{xiNLconv}) ignores an additive mode coupling term; a better model for $\xi_{\rm NL}$ \cite[see][]{rpt} sets 
\begin{equation}
 \xi_{\rm NL}(s) = \xi_{\rm L}\otimes G + \xi_{\rm MC}(s)
 \label{xiRPT}
\end{equation}
where the first term is the convolution in Eq.(\ref{xiNLconv}) and 
\begin{equation}
  \xi_{\rm MC}(s)\approx \frac{\partial\xi_{\rm L}(s)}{\partial\ln s}\,
                       \frac{\bar\xi_{\rm L}(s)}{3}
  \ {\rm where}\ 
  \frac{\bar\xi_{\rm L}(s)}{3} = \int_0^s \frac{dy}{s}\,\frac{y^2}{s^2}\,\xi_{\rm L}(y).
                       \label{xiMC}
\end{equation}
If $\xi_{\rm L}$ is given by Eq.(\ref{xiLpoly}) then $d\xi_{\rm L}/d\ln r$ is a polynomial in $ka_k (r/\sigma)^k$.  Although $\bar\xi_{\rm L}$ is also a polynomial, we should resist the temptation to use this expression because, in practice, we do not fit over the full range of $r$, so there is no guarantee that our fit works at small $r$.  Instead, we use the fact that $\bar\xi_{\rm NL}\approx\bar\xi_{\rm L}$, because the volume integral is dominated by the large scales on which linear theory should be a reasonable approximation (except around the BAO feature).  Therefore we can simply use the measured $\bar\xi_{\rm NL}$ for this term.  Hence, to include mode coupling, in Step~1 above we fit to 
\begin{equation}
  \xi_{\rm NL}(s)\approx \sum_{k=0}^n c_k \Big[ \mu_k(x)
   + kx^k \,\frac{\bar\xi_{\rm NL}(s)}{3} \Big] 
   \ {\rm with}\ x\equiv \frac{s}{\Sigma},
  \label{xiNLmc}
\end{equation}
after which we insert the fitted $c_k$ in Steps~2 and~3. 


\subsection{Biased tracers:  Scale-independent bias}
In practice, we only ever observe biased tracers of the dark matter distribution.  If the biased field is linearly proportional to the matter fluctuation field, $\delta_b = b\delta_{\rm DM}$, where $b$ is a constant, then $\xi_b(r) = b^2\xi_{\rm DM}(r)$.  In this case, because $b$ does not depend on $r$, $\xi_b$ has the same shape as $\xi_{\rm DM}$.  Hence, although the bias $b$ changes the amplitude of the correlation function, it does not change its shape.  In terms of the polynomial based description of convolution, this simply means that one determines the combination $b^2 c_k$.  Therefore, if we are ignoring the mode coupling piece when reconstructing, then we need make {\em no} change to Steps~1-3.

If we assume $\xi_{\rm NL}^b = b^2\xi_{\rm NL}^{\rm DM}$ and that $\xi_{\rm NL}^{\rm DM}$ includes mode coupling (this is the most common assumption, e.g. \cite{sanchezDR12}) then we must replace $\bar\xi_{\rm NL}\to \bar\xi_{\rm NL}/b^2$ to account for the fact that the observed $\bar\xi_{\rm NL}$ already includes a factor of $b^2$.  Since $b$ is not known \textit{a priori}, we must treat it similarly to $\Sigma$, so reconstruction will depend on both $\Sigma$ and $b$.  In practice, the importance of the mode coupling term is tracer-dependent:  e.g., Figs.~5 and~7 of \cite{bkPeaks} suggest that the mode coupling only matters for the most biased tracers.
In addition, the smearing for biased tracers differs slightly from that for dark matter \cite{sd2001}; this is sometimes called `velocity bias' \cite{rsdPeaks,baoPeaks}.

\begin{figure*}
 \centering
 \includegraphics[width=0.9\hsize]{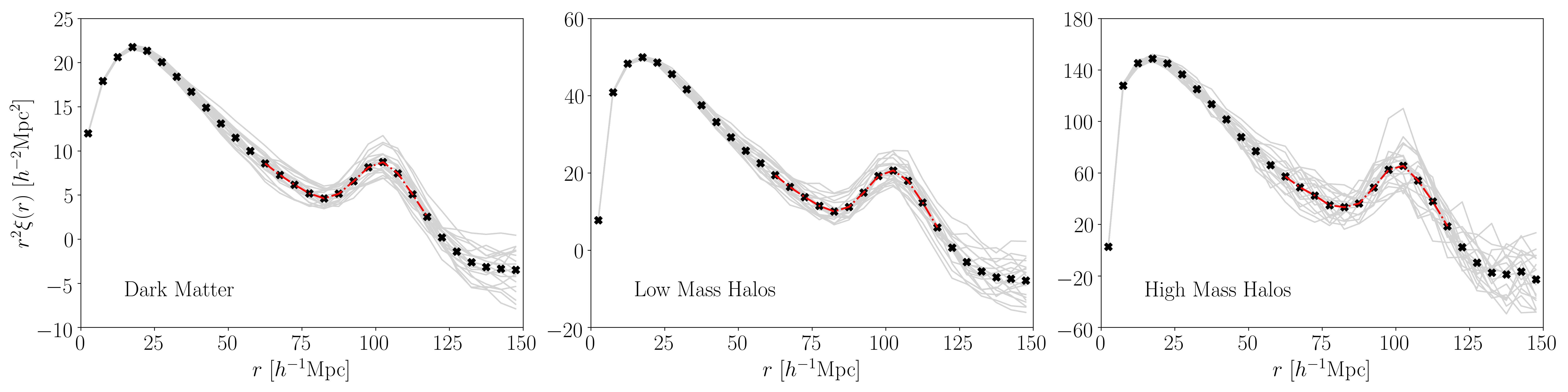}
 \caption{\label{fig:fitXi} 
   Measured correlation functions for dark matter (left), low mass halos (middle) and massive halos (right) in the $z=0.5$ outputs of the 20 simulations in our ensemble. Light grey curves in each panel show the correlation functions in each realization. The thick black curve shows the mean of the measurements and the dashed red curve shows the result of fitting Eq.(\ref{xiNLb}) with $b_{01}=0$ to it, over the range 60-120$h^{-1}$Mpc.  Note the difference in the $y$-axes:  massive halos are more strongly correlated.}
\end{figure*}

\subsection{Scale dependent bias}
The analysis is only slightly more complicated if the bias is scale-dependent.  In this case, one expects 
\begin{equation}
  \xi_{\rm L}^b \approx b_{10}^2 \xi_{\rm L}
  + 2b_{10}b_{01}R_b^2 \nabla^2\xi_{\rm L}
  + b_{01}^2R_b^4\nabla^2\nabla^2\xi_{\rm L}
  \label{xiLb}
\end{equation}
where $b_{10}$, $b_{01}$ and $R_b$ are constants \cite{bkPeaks}, so 
the evolved (smeared + mode-coupled) biased correlation function is
\begin{align}
  \xi_{\rm NL}^b(s) &\approx b_{10}^2 \sum_{k=0}^n c_k\,\mu_k \nonumber\\
  & \quad + 2b_{10}b_{01}(R_b/s)^2 \sum_k c_k\,\Big(2s\mu_k^{(1)} + s^2\mu_k^{(2)}\Big)\nonumber\\
  & \quad + b_{01}^2 (R_b/s)^4 \sum_k c_k\,\Big(4s^3\mu_k^{(3)} + s^4\mu_k^{(4)}\Big)\nonumber\\
  & \quad + \frac{\partial\xi_{\rm L}^b}{\partial\ln s} \frac{\bar\xi_{\rm NL}^b(s)}{3b_{10}^2},
  \label{xiNLb}
\end{align}
where $\mu_k^{(n)}\equiv d^n\mu_k/ds^n$ and $\partial\xi_{\rm L}^b/\partial\ln s$ in the final (mode-coupling) term can also be written in terms of the $c_k$.  Thus, scale-dependent bias simply complicates the functions that multiply the $c_k$ coefficients.


\begin{table*}
\begin{center}
  \begin{tabular}{c|c|c|c|c|c}
     \hline
    Tracer & $b_{10}$ & $r_{\rm LP-pre}$ & $r_{\rm LP-rec}$ & $r_{\rm infl-pre}$ & $r_{\rm infl-rec}$\\ 
     \hline
      DM &  1  & $92.43\pm 0.24$ & $92.98\pm 0.21$ & $92.78\pm 0.26$ & $93.42\pm 0.22$ \\
      LM & 1.3 & $92.24\pm 0.27$ & $93.06\pm 0.22$ & $92.57\pm 0.28$ & $93.35\pm 0.24$ \\
      HM & 2.6 & $92.06\pm 0.46$ & $92.97\pm 0.39$ & $92.45\pm 0.49$ & $93.49\pm 0.41$ \\
     \hline
    \end{tabular}
\end{center}
\caption{Linear point and inflection scales (in $h^{-1}$Mpc) in the pre- and post-reconstruction correlation functions, estimated by fitting 9th-order Laguerre-based functions to the $z=0.5$ two-point correlation functions (bins of width $3h^{-1}$Mpc over the range 60-120$h^{-1}$Mpc) of dark matter, low mass halos and high mass halos in an effective comoving volume of nearly 27~$h^{-3}$Gpc$^3$. Laguerre reconstruction brings $r_{\rm LP}$ and $r_{\rm infl}$ closer to their linear theory values  without inflating the errors.}
  \label{tab:rLPs}
\end{table*}

\begin{figure*}
 \centering
 \includegraphics[width=0.95\hsize]{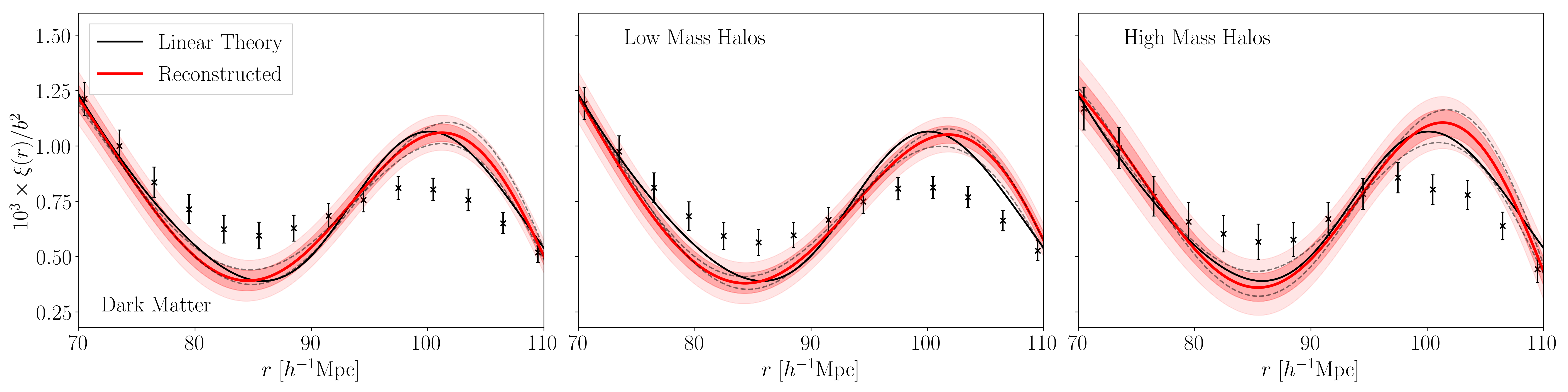}
 \caption{\label{fig:shape} 
   Linear theory correlation functions reconstructed from the fits to the nonlinear correlations shown in the previous figure.  Red curve shows the reconstruction when using: the fiducial value of smearing, the correct value of $b_{10}$ when modeling the mode coupling term, and no correction for scale dependent bias (i.e. $b_{01}=0$).  Pink bands show the result of propagating the 1- and 2-standard deviation uncertainties on the fits to the reconstruction, as described in the main text.  Dashed grey curves show reconstructions when the smearing is assumed to be larger or smaller by 10\%.  Black curve, same in each panel, shows the actual linear theory shape.  This shape is quite well reconstructed, especially in between the peak and dip scales.} 
\end{figure*}

Eq.(\ref{xiNLb}) is the most general expression that we use in Step~1 of our reconstruction algorithm.  It illustrates the three bits of prior information about the background or fiducial cosmology that are needed as one makes the reconstruction ever more sophisticated.  To undo smearing, one only needs $\Sigma$; to include mode coupling as well, one must know the constant bias parameter $b_{10}$; and if the bias is scale dependent, then one additionally needs the combination $(b_{01}/b_{10}) (R_b/\Sigma)^2$.  We generically expect $R_b/\Sigma\sim 1$, so this combination is large if $b_{01}/b_{10}\gg 1$.  Although there is some physical understanding of, e.g., how this ratio depends on halo mass \cite{bk2dc,bkConsistency}, for all the tests that follow, we always set $b_{01}=0$.  


\section{Results}\label{sec:results}
We validate our methodology using the dark matter and halo distributions at $z=0.5$ in the {\small ABACUS} simulation suite \cite{abacus}, which provides 20 periodic boxes each of comoving size $1100h^{-1}$Mpc -- an effective comoving volume of nearly $27 h^{-3}$Gpc$^3$ -- in which the background cosmology is a flat $\Lambda$CDM model with $(\Omega_{\rm cdm}h^2,\Omega_b h^2) = (0.1199,0.02222)$, and $(h,n_s,\sigma_8) = (0.6726,0.9652,0.83)$.  The associated values of $r_{\rm LP}$, $r_{\rm infl}$ and $\Sigma$ are 93, 93.4, and 4.6$h^{-1}$Mpc.  

Ref.\cite{abacus} also provide a suite of 16 additional simulations having the same cosmological parameters, but with a different treatment of the small-scale physics.  We refer to these as the Emulator runs, and discuss our analyses of these runs in Appendix~\ref{sec:AE}.  Since the 20 {\small ABACUS} runs are expected to be more reliable \cite{abacus}, we only present results for them in the main text.

\subsection{Initial estimates} 
The symbols in Fig.~\ref{fig:fitXi} show correlation functions measured in bins that are $3h^{-1}$Mpc wide for dark matter (left), halos more massive than $8\times 10^{11}h^{-1}M_\odot$ (middle), and halos more massive than $3\times 10^{13}h^{-1}M_\odot$ (right) in the $z=0.5$ outputs.  We will sometimes refer to these as the DM, LM and HM samples. The number densities of these three types of tracers are $6.9\times 10^{-3}$(Mpc$^{-1} h$)$^3$, $5.5\times 10^{-3}$(Mpc$^{-1} h$)$^3$, and $8.6\times 10^{-5}$(Mpc$^{-1} h$)$^3$ respectively. The halo samples have large-scale bias factors -- measured from the amplitude of their power spectra at $k<0.05h^{-1}$Mpc -- of $b_{10}=1.3$ and $2.6$.  The less biased sample is similar to that considered in \cite{PaperI}, whereas the more massive sample is similar to that which hosts the Luminous Red Galaxies used for BAO measurements. 

The dashed lines show the best fits of Eq.(\ref{xiNLb}) with $n=9$ to the mean curve traced out by these measured $\xi$.
We fit to the correlation function in $3h^{-1}$Mpc bins over the range 60-120$h^{-1}$Mpc, and use the analytic estimate of the covariance which is described in \cite{eppur, LPnus} when fitting.  (Our results are unchanged if we use the noisier covariance matrix measured directly from the 20 simulations.)  Appendix~\ref{fitDetails} illustrates how the goodness of fit (e.g. $\chi^2$/d.o.f.) varies with different choices for the order of the polynomial and bin size. 
It also shows that the $r_{\rm LP}$ values estimated from these fits are robust to reasonable changes in these choices.

The fits in Fig.~\ref{fig:shape} all have $\chi^2$/d.o.f. $\approx 1$, so using the fitted parameters $c_k$ is meaningful.  From these fits, we determine where $\xi_{\rm NL}'=0$ and $\xi_{\rm NL}''=0$, and hence find $r_{\rm LP}$ and $r_{\rm infl}$.  Table~\ref{tab:rLPs} shows that they are always smaller than the linear theory value of $93h^{-1}$Mpc, with the largest discrepancy for the most biased tracers.  
More biased tracers tend to be more massive: they assemble their mass from larger scales and have larger streaming motions.  The former potentially increases the effective smearing scale, and the latter potentially modifies the mode-coupling term as well, so mass/bias dependent shifts from linear theory are plausible.  However, with the exception of peaks-theory based models \cite{sd2001,rsdPeaks,baoPeaks} there is currently no first principles derivation of this mass dependence.

\subsection{Deconvolved/reconstructed estimates}
For the reconstruction results which follow, we set $b_{01}=0$, and we used the correct value of $b_{10}$ for the mode-coupling piece (we show results using the incorrect value shortly).  The red curves in the three panels of Fig.~\ref{fig:shape} show the result of inserting the $c_k$ obtained from fitting the dashed curves in Fig.~\ref{fig:fitXi} into Eq.(\ref{xiLpoly}), and setting the smearing scale $\Sigma$ to the fiducial value.  Propagating the errors on the fitted $c_k$ to the $a_k$ used in Eq.(\ref{xiLpoly}) yields the pink bands (which show the 1- and 2$\sigma$ uncertainties).  The solid black curve, same in all the panels, shows the linear correlation function.  Our reconstructed shape is obviously much closer to linear theory than are the original measurements, although it tends to push the peak to larger and the dip to smaller scales.  Nevertheless, Table~\ref{tab:rLPs} shows that the $r_{\rm LP}$ and $r_{\rm infl}$ scales in the reconstructed correlation functions are considerably closer to their linear theory values, and the trend with mass has been removed.  Note in addition that the reconstruction procedure does not increase the uncertainty on the inferred scales. 

\begin{figure}
 \centering
 \includegraphics[width=0.9\hsize]{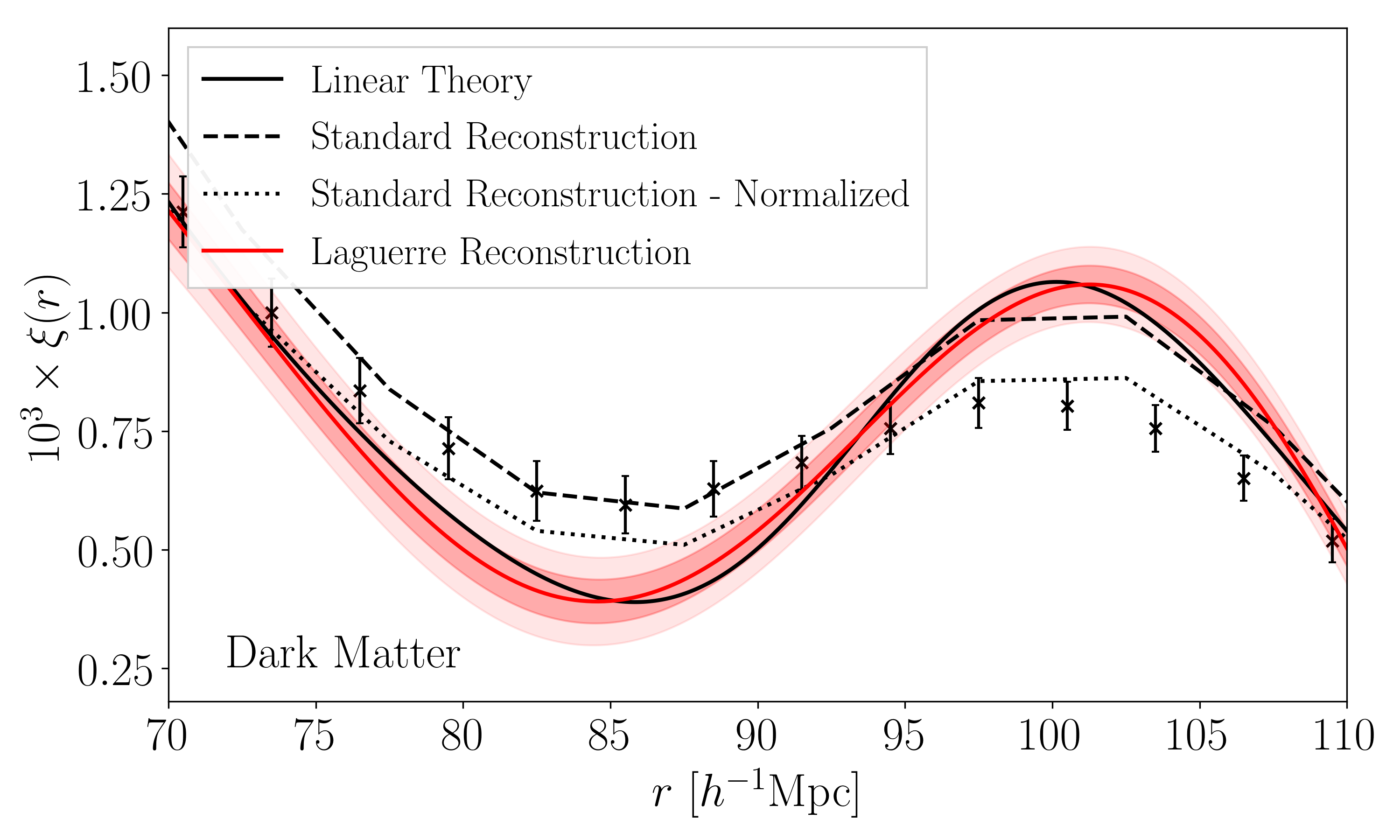}
 \caption{\label{fig:std} 
   Comparison of our Laguerre reconstruction of the shape of the dark matter correlation function with a more traditional reconstruction from Ref.\cite{baoHOD}:  dashed curve shows their `standard' reconstruction, and dotted curve shows the result of normalizing it to have the same value as linear theory at $70h^{-1}$Mpc.}  
\end{figure}

\begin{figure*}
 \centering
 \includegraphics[width=0.95\hsize]{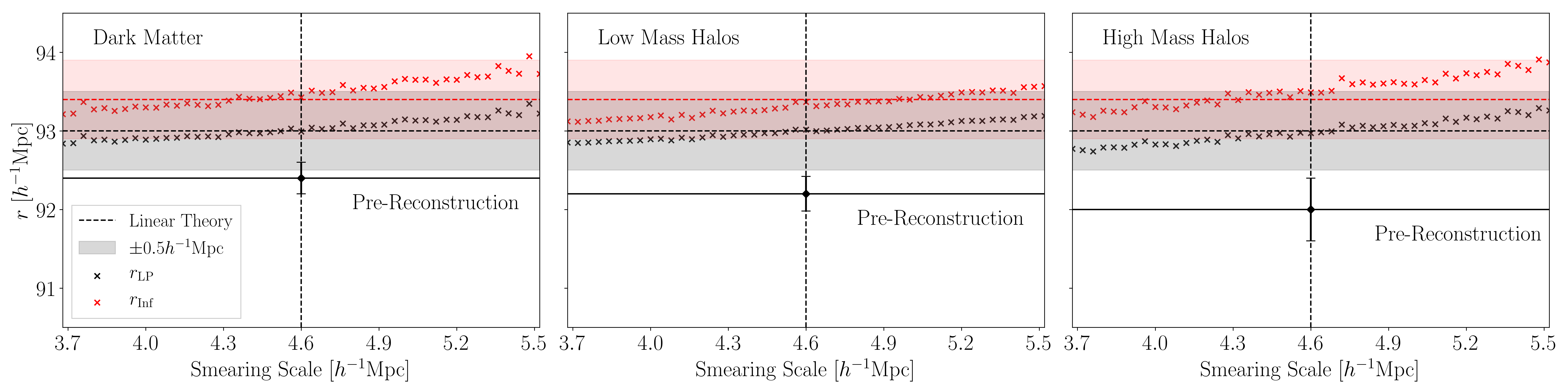}
 \caption{\label{fig:recon} 
   Degeneracy between assumed smearing scale and $r_{\rm LP}$ and $r_{\rm inf}$ in the reconstructed correlation functions, obtained from the Laguerre-fits to the symbols shown in the previous Figure.  Solid line close to the bottom of each panel shows the linear point measured in Laguerre fits, and crosses show the values in the associated reconstructions. The horizontal dashed lines show $r_{\rm LP}$ and $r_{\rm inf}$ in the linear theory $\xi_{\rm L}$. The agreement shows that our algorithm provides estimates of the distance scale that are robust to expected uncertainties in the smearing scale.}
\end{figure*}

\subsection{Comparison with standard reconstruction}
We close this section with a direct comparison of our Laguerre reconstruction with a more traditional algorithm.  For this, we have used what Ref.\cite{baoHOD} refer to as the `standard' reconstruction of the dark matter signal for these same 20 {\small ABACUS} simulations. (Similar results for the LM and HM samples are not available.)

In Fig.~\ref{fig:std}, the smooth black curve shows linear theory, symbols with error bars show $\xi_{\rm NL}$ and red curve surrounded by pink bands shows our Laguerre reconstruction $\xi_{\rm Lag}$ (same as left hand panel of Fig.~\ref{fig:shape}).  The dashed curve is from Ref.\cite{baoHOD} (provided in $5h^{-1}$Mpc bins), and the dotted curve shows the result of normalizing it to have the same value as linear theory at $70h^{-1}$Mpc. These show the correlation function measured on the reconstructed density field.  It is apparent that our simpler Laguerre-based reconstruction is closer to the linear theory shape over a wider range of scales.

However, what really matters is the distance scale that one estimates from these (dashed or dotted) curves.  The `standard' procedure involves fitting a $\Lambda$CDM template to the (dashed or dotted) curves.  Instead, we will treat them similarly to how we treat $\xi_{\rm Lag}$.  Namely, we fit a 9th-order simple polynomial to the dotted curve.  Although this has $\chi^2$/d.o.f. = 9.4, indicating a bad fit, the associated $r_{\rm LP}$ is $92.86\pm 0.32 h^{-1}$Mpc.  This is a $\sim 0.5\%$ improvement on $r_{\rm LP-pre}$ (c.f. Table~\ref{tab:rLPs}), even though $r_{\rm LP}$ was not used to calibrate this `standard' reconstruction algorithm.

Although $r_{\rm LP-rec}$ from our simpler Laguerre-based reconstruction is slightly more accurate (Table~\ref{tab:rLPs}), the peak and dip positions in $\xi_{\rm Lag}$ are slightly shifted in opposite directions with respect to linear theory.  These shifts nearly cancel out for $r_{\rm LP}$, but may have a greater impact on more traditional estimators of the distance scale.  Leveraging the improved Laguerre-reconstructed shape for other distance scale estimators is interesting, but beyond the scope of this work.  

Finally, we note that the CPU time and memory of all of the more traditional reconstruction algorithms \cite{recPW, recIterate, recHE, eFAM2019, royaMAK} increases with the number of objects (in the simulation or survey), in some cases dramatically.  In contrast, since Laguerre reconstruction boils down to fitting a curve to the measured correlation function, the associated computational time scales with the number of bins (as opposed to number of particles).  Therefore, CPU time/memory requirements are miniscule.

\section{Realistic constraints}\label{sec:stats}
Both the Laguerre and `standard' density field reconstructions depend on input parameters.  E.g., $\xi_{\rm Lag}$ depends on an assumed smoothing scale $\Sigma$ and, if one wants to account for mode-coupling, a bias factor $b_{10}$.  (Accounting crudely for scale-dependent bias would require one additional parameter, $b_{01}$.)  Likewise, `standard' reconstruction assumes a fiducial cosmology and bias prescription.  The previous section (Table~\ref{tab:rLPs} and~Fig.~\ref{fig:std}) showed that both work well if the fiducial choice is good: for $\xi_{\rm Lag}$, this means we used the correct $\Sigma$ and $b_{10}$ and simply set $b_{01}=0$.

In real datasets, the appropriate $\Sigma$ and $b_{10}$ to use are not known perfectly.  Accounting for this will almost certainly increase the error bars in Table~\ref{tab:rLPs}, and may even bias the $r_{\rm LP}$ values, for both the Laguerre and `standard' reconstructions.  This raises the question of how to incorporate such systematic uncertainties on the reconstruction in a principled way.  In the Laguerre context, this is straightforward:  We first study the dependence on $\Sigma$, and then on both $\Sigma$ and $b_{10}$.  

\begin{figure}
 \centering
 \includegraphics[width=0.95\hsize]{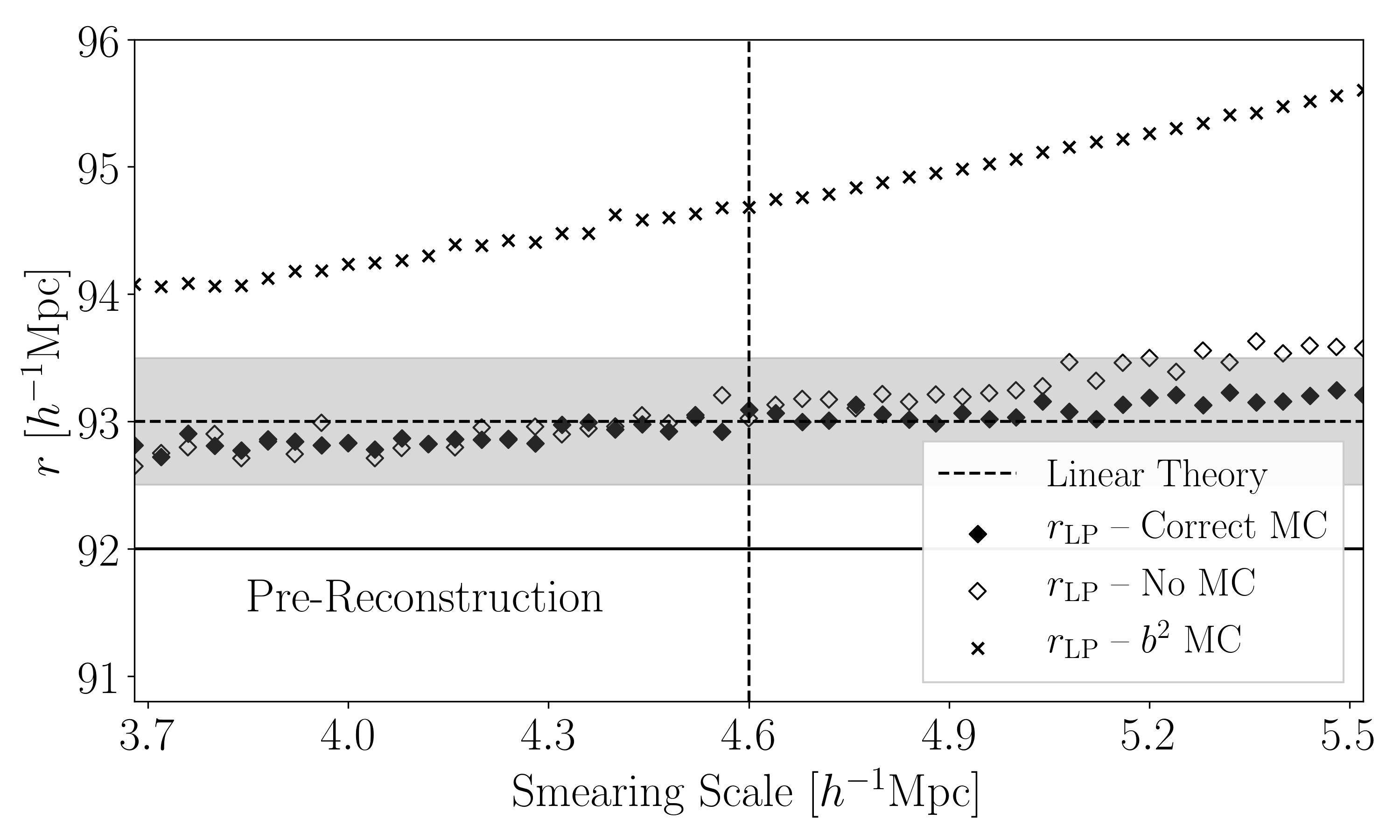}
 \caption{\label{fig:noMC} 
   Linear point for massive halos ($b=2.6$) pre- (solid line) and post-reconstruction (filled symbols) when we ignore mode-coupling altogether (open symbols), or we overestimate its value by a factor of $b_{10}^2$ (crosses).}
\end{figure}

\subsection{Dependence on assumed smearing scale}

\begin{figure*}
 \centering
 \includegraphics[width=\hsize]{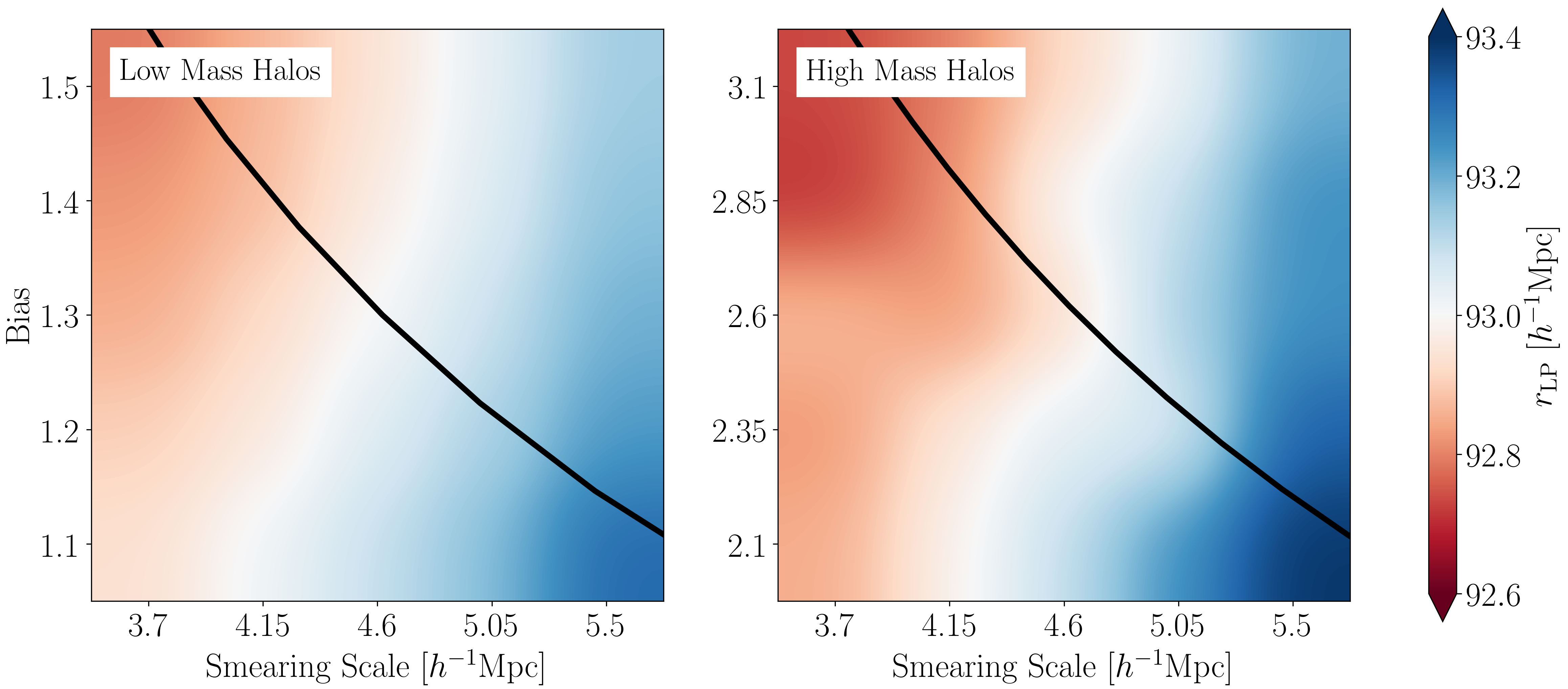}
 \caption{\label{fig:ba}
   Dependence of $r_{\rm LP}$ in the reconstructed correlation function on assumed values of bias and smearing scale, if we ignore scale dependence of bias. Black curve in each panel shows $b$ vs $\Sigma_{\rm fid}/(b/b_{\rm fid})$, the locus along which one should read off $r_{\rm LP}$ values so as to get more realistic uncertainties on $r_{\rm LP}$.  We set $\Sigma_{\rm fid}=4.6h^{-1}$Mpc, $b_{\rm fid}=1.3$ (left) and 2.6 (right).  If $\Sigma_{\rm fid}$ is also unknown then this will shift the curves to the left or right, potentially broadening the error estimate further.}  
\end{figure*}

For Laguerre reconstruction, the assumed smearing scale affects the transformation from $c_k$ to $a_k$ (Eq.~\ref{xiNLpoly}).  The dashed grey curves in Fig.~\ref{fig:shape} show the result of changing the smearing scale by $\pm 10\%$:  larger $\Sigma$ results in a reconstructed $\xi$ that is more sharply peaked.  Fig.~\ref{fig:recon} explores this further for dark matter (left) as well as low and high mass halos (middle and right).  The bar along the bottom of each panel shows $r_{\rm LP-pre}$ of Table~\ref{tab:rLPs}: the linear point estimated from the nonlinear correlation function (i.e. where $\xi_{\rm NL}'=0$ and $\xi_{\rm NL}''=0$ in Fig.~\ref{fig:fitXi}).  The vertical dashed line --- same in all three panels --- shows the expected smearing scale for the dark matter.  This is the value one would use as the fiducial smearing. The symbols show how the $r_{\rm LP}$ and $r_{\rm infl}$ values from the corresponding reconstructed $\xi_{\rm Lag}$ depend on the assumed smearing scale.  If one over-estimates the smearing, then one `reconstructs' too much, so $r_{\rm LP}$ in the reconstruction is pushed to larger scales.  However, this is a small effect:  varying our guess for the smearing scale by $\pm 20\%$ relative to the fiducial value only changes the reconstructed values by $\pm 0.5\%$.  As uncertainties on the amount of smearing are smaller than this, Fig.~\ref{fig:recon} shows that our algorithm provides a simple and robust method of reconstructing the distance scale that only depends weakly on the assumed background model.  (In practice, one would marginalize over a prior distribution of $\Sigma$ values that would be survey specific.)

\subsection{Dependence on smearing scale and halo bias}\label{sec:ba}
The impact of $b_{10}$ -- which affects the strength of our correction for mode-coupling -- is also straightforward to assess.  The crosses in Fig.~\ref{fig:noMC} show the result of including the mode-coupling term but not dividing by the factor of $b_{10}^2$, so that the strength of this term is over-estimated, for the high mass halo sample.  This pushes $r_{\rm LP}$ in the reconstructions to too high values.  The open solid circles show the other extreme in which the mode-coupling term is omitted altogether. Evidently, accounting for mode-coupling matters little. This is attractive, since ignoring mode coupling allows one to be more agnostic about the underlying model.


Finally, Fig.~\ref{fig:ba} illustrates how $r_{\rm LP}$ in the reconstructions depends on both bias and smearing scale (results for $r_{\rm infl}$ are similar).  In both panels, the correct value $r_{\rm LP}=93h^{-1}$Mpc can be recovered along the white region approximately defined by $b_{10}-b_{\rm true} \approx 0.7 (\Sigma/[h^{-1}{\rm Mpc}] - 4.6)$.  Note that the color scheme we have chosen shows variations in $r_{\rm LP}$ of $\pm 0.5\%$ around the fiducial value.  Evidently, $20\%$ misestimates of the bias and smearing scale only affect $r_{\rm LP}$ at the $0.5\%$ level.

In practice, one would quantify the effects of such systematics on the accuracy and precision of the distance scale estimate by marginalizing over some prior distribution of $\Sigma$ and $b_{10}$ values. The priors are likely to be correlated.  E.g., the clustering strength is proportional to $b\sigma_8$, whereas the smearing scale is proportional to $\sigma_8$.  Since $b = (b\sigma_8)/\sigma_8\propto \sigma_8^{-1}$ whereas $\Sigma\propto \sigma_8$, one might expect realistic uncertainties on $r_{\rm LP}$ to be associated with averaging along a curve, $b\propto \Sigma^{-1}$, in the $b-\Sigma$ plane.

The `observers\textquotesingle' version of the `theorists\textquotesingle' discussion above is as follows. Suppose one used the observed $P_{\rm obs}(k)$ to estimate a smearing scale $\Sigma_b$.  This will be wrong because $P_{\rm obs}$ carries bias factors (hence the subscript $b$) and is nonlinear, whereas the actual smearing scale $\Sigma$ should use $P_{\rm Lin}(k)$ of the dark matter.  Since the integral which defines $\Sigma$ down-weights nonlinear scales (by a factor of $1/k^2$), the nonlinear value should not be too different from that in linear theory, so we expect $\Sigma = \Sigma_b/b$ to be a reasonable approximation.  This makes $b$ -- the same parameter which affects the normalization of the mode-coupling contribution -- the only unknown.  As a result, the two dimensional plane of unknown parameters ($b$ vs $\Sigma$) becomes a one-dimensional curve:  $b/b_{\rm fid} = (\Sigma_b/b)$.

\begin{figure}
 \centering
 \includegraphics[width=0.95\hsize]{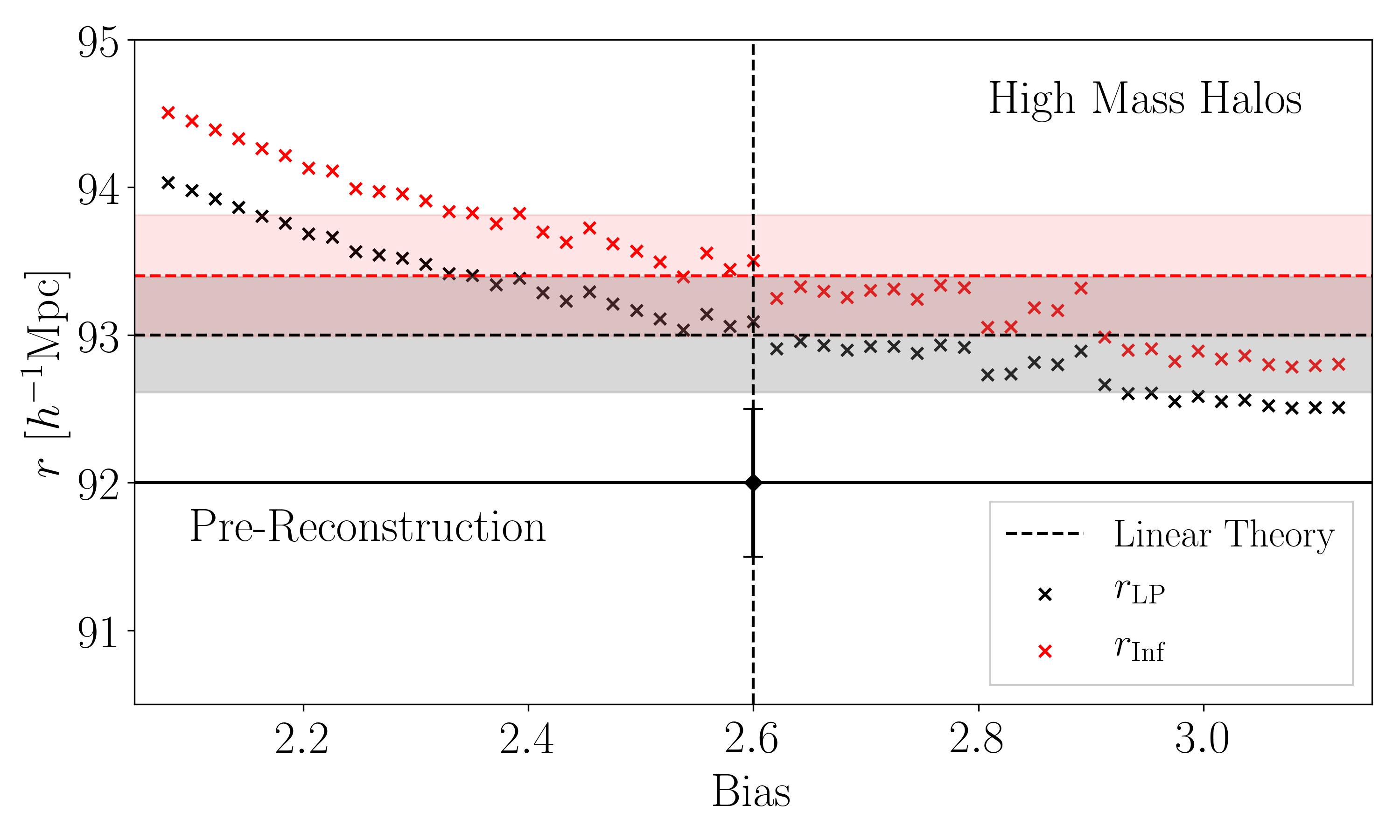}
 \caption{\label{fig:b_rLP}
 Degeneracy between assumed bias factor $b_{10}$ and $r_{\rm LP}$ and $r_{\rm inf}$ in the reconstructed correlation functions, as one moves along the black curve shown in right hand panel of Figure~\ref{fig:ba}.   Pink and grey bands show the uncertainties quoted in Table~\ref{tab:rLPs}, which assume that $b_{10}$ and $\Sigma$ are known perfectly:  accounting for the fact that they are not broadens the uncertainty on the distance scale. }
\vspace{-0.5cm}
\end{figure}

The thick black curve in each panel of Figure~\ref{fig:ba} shows $b = \Sigma_{\rm fid}/(b/b_{\rm fid})$:  this is the direction along which one should read-off $r_{\rm LP}$ values so as to get more realistic error bars, if $\Sigma_b/b$ is indeed equal to $\Sigma_{\rm fid}=4.6h^{-1}$Mpc when $b$ is equal to the correct value $b_{\rm fid}=b_{10}$.  To put it another way, if one has a given range of input smearing scale or bias in mind, one can bracket the uncertainty this would produce in $r_{\rm LP}$ by reading off the black curves. The symbols in Figure~\ref{fig:b_rLP} show the results of this exercise for the massive halo sample.  They show how $r_{\rm LP}$ and $r_{\rm inf}$ in the reconstructed $\xi_{\rm Lag}$ change as one moves along the thick black curve shown in the right hand panel of Figure~\ref{fig:ba}.  The colored bands show the error bars in Table~\ref{tab:rLPs} which assume $b$ (and $\Sigma$) are known perfectly.  Some symbols lie outside these bands illustrating how accounting for uncertainties in the parameters used to reconstruct can broaden the errors on the inferred distance scale.

If $\Sigma_b/b_{\rm fid}$ does not equal $\Sigma_{\rm fid}$ exactly, then this will shift the thick black curve (to the left or right) in the $b-\Sigma$ plane.  Therefore, allowing for uncertainties in the $\Sigma_b/b_{\rm fid}\approx\Sigma_{\rm fid}$ assumption will further degrade the constraints. While this shifting and associated degradation will be survey-specific, because lines of fixed $r_{\rm LP}$ run approximately perpendicular to the black curves in each panel, the degradation in constraining power may not be crippling.  Thus, although assuming perfect knowledge of the input parameters required for reconstruction (whether Laguerre or full density field) leads to underestimates of the true uncertainties on the inferred distance scale, at least for Laguerre reconstruction, making more realistic estimates is straightforward.


\subsection{Relation to previous LP analyses}\label{sec:compare}
Before ending this section, it is worth contrasting our methodology with previous LP analyses \cite{PaperI,PRDmocks,PRLboss,LPnus}, which fit $\xi_{\rm NL}$ to a simple polynomial and then multiply the $r_{\rm LP-pre}$ derived from it by a factor of $1+\epsilon$ with $\epsilon=0.005$.  For the discussion which follows, it is useful to distinguish between the step which multiplies $r_{\rm LP-pre}$ by $1+\epsilon$ and the decision to set $\epsilon=0.005$.  

We begin by noting that both the simple-polynomial and our current Laguerre-based methods are motivated by the fact that Eq.~(\ref{xiNLconv}) is a good approximation.  Next, we note that there is no \textit{a priori} reason for fitting a simple polynomial to $\xi_{\rm NL}$.  
Hence, our Step~1 which fits to Laguerre functions is essentially no different from previous LP-related work.  What {\em is} different is that we have a reason for fitting with Laguerres rather than simple polynomials.  However, regardless of motivation, the estimate of $r_{\rm LP}$ which results from this choice should not -- and we have checked that it does not -- depend on what family of curves we choose to fit (provided they return acceptable fits).  In this respect, both our methodology and the LP approach are agnostic about the (in principle unknown) shape of the dark matter correlation function.  We turn therefore to Steps 2 and 3 of Laguerre reconstruction.

In effect, the factor of $1+\epsilon$ in previous LP work is a crude way of correcting for the fact that $r_{\rm LP}$ in $\xi_{\rm NL}$ differs from that in $\xi_{\rm L}$ because the shape of $\xi_{\rm NL}$ differs from that of $\xi_{\rm L}$.  In this respect, its goal is to undo the effects of the convolution in Eq.~(\ref{xiNLconv}) (illustrated in Figure~\ref{fig:smoothing}), and whatever else causes the shapes of $\xi_{\rm NL}$ and $\xi_{\rm L}$ to differ.  The goal of Steps 2 and 3 in our algorithm here is analogous.  The assumption that the convolution is with a Gaussian singles out Laguerre functions because they are the ones for which the deconvolution problem is trivial.  So, by using Laguerres, we make more explicit use of the Gaussian assumption than previous LP work.

The only remaining question is what to use for $\Sigma$ when deconvolving, and this is analagous to choosing a value of $\epsilon$, both conceptually, and statistically (because, once $\Sigma$ is fixed, the formal uncertainty on $r_{\rm LP}$ both pre- and post-reconstruction is unchanged).  While this connection between $\epsilon$ and $\Sigma$ is not necessary for the LP approach, by tying $\Sigma$ to the Gaussian convolution kernel, our Laguerre reconstructions provide some intuition into what $\epsilon$ means, at least in the context of $\Lambda$CDM models.

Perhaps the only real difference between Laguerre reconstruction and multiplication by a corresponding $1+\epsilon$ is that if the evolved correlation function does {\em not} show a peak or a dip, then the usual LP approach cannot estimate $r_{\rm LP}$.  However, even if the Laguerre fit to $\xi_{\rm NL}$ does not show a peak or dip, the reconstructed $\xi_{\rm L}$ may, so a distance scale estimate may still be possible.

Nothing in the discussion so far singles out the value $\epsilon=0.005$ as being special.  This choice was calibrated by \cite{PaperI} from a set of $\Lambda$CDM simulations with CMB-motivated values of the cosmological parameters, and $\sigma_8\sim 0.8$ at $z=0$, because it provided a corrected $r_{\rm LP}$ value that was within $0.5\%$ of the linear theory value at all $z$.  Fig.2 of \cite{LPruler} shows that $\epsilon=0.005$ works well -- in the sense that it corrects $r_{\rm LP}$ to within $0.5\%$ of the linear theory value -- for a wide range of cosmological parameters.  Indeed, multiplying the $r_{\rm LP-pre}$ values in Table~\ref{tab:rLPs} by 1.005 does bring them to within $0.5\%$ of linear theory (although the systematic trend with halo mass remains).

Since $\Sigma$ depends on cosmology and redshift, the correspondence between $\epsilon$ and $\Sigma$ in the preceding paragraphs shows that the choice $\epsilon=0.005$ corresponds to a crude marginalization over the interesting range of $\Sigma$ values, with the associated degradation in precision yielding a systematic uncertainty of $0.5\%$.  And indeed, as Fig.~\ref{fig:recon} shows, a $0.5\%$ systematic arising from uncertainties on the correct value of $\Sigma$ is reasonable.  In effect, marginalizing over $\Sigma$ and $b_{10}$ values in Fig.~\ref{fig:ba} allows one to make a slightly more careful estimate of the distance scale and its uncertainties.


\section{Discussion}\label{sec:disc}
On BAO scales, the relation between the linear theory correlation function $\xi_{\rm L}$ and the biased and nonlinearly evolved $\xi_{\rm NL}^b$ is understood to be quite well approximated by the sum of a convolution term and a `mode-coupling' term (Eqs.~\ref{xiNLconv}, \ref{xiRPT} and~\ref{xiMC}).  We show that if $\xi_{\rm L}$ can be approximated by a polynomial (Eq.~\ref{xiLpoly}), then $\xi_{\rm NL}^b$ can be written analytically using associated Laguerre functions (Eqs.~\ref{xiNLpoly}, \ref{xiNLmc} and~\ref{xiNLb}).  This motivates a three-step algorithm (Section~II.B) which approximately reconstructs the original shape of $\xi_{\rm L}$ from the measured one (Figs.~\ref{fig:fitXi} and~\ref{fig:shape}).  We use the linear point scale, $r_{\rm LP}$ of Eq.~(\ref{eq:rLP}), to quantify the accuracy and precision of the reconstruction.

Each step of our algorithm uses some prior information about the background cosmology:  depending on the desired level of sophistication, a smearing scale, constant bias factor, and scale dependent bias factor must be assumed (Eq. \ref{xiNLb} and related discussion).  Our tests indicate that, for a wide variety of tracers, only the smoothing scale is required (Fig.~\ref{fig:noMC} and related discussion).  If the required prior information is known precisely, then our algorithm recovers $r_{\rm LP}$ to subpercent precision, even for highly biased tracers (Fig.~\ref{fig:recon} and Table~\ref{tab:rLPs}).

In practice, the required prior information is not known perfectly.  We show that the $r_{\rm LP}$ estimated from the Laguerre reconstructed correlation function is not strongly dependent on the assumed values:  20\% variations in the smearing scale and bias factor change $r_{\rm LP}$ by less than $0.5\%$ (Figs.~\ref{fig:recon} and~\ref{fig:ba}).  Our analysis shows how to include such systematic uncertainties when quantifying the precision of the distance scale estimate, with minimal assumptions about the background cosmology or the nature of the bias of the observed tracers (Fig.~\ref{fig:b_rLP} and associated discussion).


As the prior information which our Laguerre reconstructions require is similar to that used by more traditional reconstruction algorithms \cite{recPW, recIterate, recHE, eFAM2019, royaMAK}, our methodology provides a simple, cheap and accurate sanity check of these more elaborate and computationally expensive schemes. A direct comparison of the shape we reconstruct with that returned by one of these more traditional algorithms is encouraging (Fig.~\ref{fig:std}).  In future work, we intend to explore the synergies between our Laguerre reconstructions of the correlation function shape and more traditional estimates of the BAO distance scale.  For instance, Laguerre reconstruction provides a straightforward way of estimating the degradation in constraining power which results when the parameters on which reconstruction depends are not perfectly well known (Figs.~\ref{fig:recon} and~\ref{fig:ba} and associated discussion).  


Although our tests were performed using distances that were not perturbed by redshift space distortions, they should apply essentially without change to the redshift space monopole (the smearing scale and bias factors will be slightly modified, but the overall structure will not).
This is the subject of work in progress.  In the meantime, as our algorithm is simple, computationally cheap and accurate, we hope it will be useful in next generation BAO datasets.  

Finally, although all our analysis used correlation functions which were estimated in bins, our results suggest useful synergy with recent `least squares' estimators which do not require binning \cite{lsqXi18,nobinXi21}.  These expand the correlation function in a set of basis functions, and our work shows that generalized half-integer Laguerre functions are a particularly interesting choice for BAO studies.  We intend to explore this synergy in future work.


\bigskip

\begin{acknowledgments} 
  We are grateful to the referee for a detailed report, to Y. Duan for providing the `standard' reconstructions that we show in Figs.~\ref{fig:std} and~\ref{fig:stdAE}, to S. Anselmi for advice on estimating the LP prior to reconstruction and noting that it is important to consider the Abacus and Emulator runs separately, and to L. Garrison, G. Parimbelli and G. Starkman for helpful discussions.  FN and RKS thank the Munich Institute for Astro- and Particle Physics (MIAPP) which is funded by the Deutsche Forschungsgemeinschaft (DFG, German Research Foundation) under Germany's Excellence Strategy – EXC-2094 – 390783311, for its hospitality during the summer of 2019.
  FN acknowledges support from the National Science Foundation Graduate Research Fellowship (NSF GRFP) under Grant No. DGE-1845298.
 IZ acknowledges support from NSF grant AST-1612085.
\end{acknowledgments}

\bibliography{LP_recon_rev2}

\appendix

\section{Generalized Laguerre functions}\label{sec:Lab}
Eq.\ref{chi3-moments} of the main text uses half-integer generalized Laguerre functions.  We describe some of their relevant properties below.  

\subsection{Explicit expressions}\label{sec:explicit}
Starting from $L_0^{(1/2)} = 1$, $L_1^{(1/2)}=(x^2+3)/2$,
\begin{align}
  L_{-1/2}^{(1/2)}(-x^2/2) &= \sqrt{\frac{2}{\pi}}\frac{{\rm erf}(x/\sqrt{2})}{x}
                          \qquad{\rm and}\\
  L_{1/2}^{(1/2)}(-x^2/2) &=  (x^2 + 1)\, L_{-1/2}^{(1/2)}(-x^2/2)  
                            + \frac{{\rm e}^{-x^2/2}}{\pi/2}, \nonumber
\end{align}
the others can be generated from 
\begin{equation}
 \beta L_{\beta}^{(\alpha)}(z) = (\alpha + 2\beta -1-z)\,L_{\beta-1}^{(\alpha)}(z) - (\alpha + \beta -1)\,L_{\beta-2}^{(\alpha)}(z).
\end{equation}
Thus, the $\mu_k$ of Eq.(\ref{chi3-moments}) are
\begin{align}
	\mu_1(x) &= (x + 1/x)\,E_1(x) + E_2(x) \nn\\
	\mu_2(x) &= 3 + x^2\nn\\
	\mu_3(x) &= (x^3 + 6x + 3/x)\,E_1(x) + (x^2 + 5)\, E_2(x)\nn\\
	\mu_4(x) &= x^4 + 10x^2 + 15 \nn\\
	\mu_5(x) &= (x^5 + 15x^3 + 45x + 15/x)\,E_1(x)\nn \\
			 &\quad + (x^2 + 3)(x^2 + 11)\,E_2(x) \\
	\mu_6(x) &= x^6 + 21x^4 + 105x^2 + 105 \nn\\
	\mu_7(x) &= (x^7 + 28x^5 + 210x^3 + 420x + 105/x)\,E_1(x) \nn \\
	                 & \quad + (x^6 + 27x^4 + 185x^2 + 279)\,E_2(x)\nn\\
	\mu_8(x) &= x^8 + 36x^6 + 378x^4 + 1260x^2 + 945\nn\\ 
	\mu_9(x) &= (x^9 + 45x^7 + 630x^5 + 3150x^3 + 4725x + 945/x)\,E_1(x)\nn\\
                 & \quad + (x^8 + 44x^6 + 588x^4 + 2640x^2 + 2895)\,E_2(x),\nn
\end{align}
where $E_1(x)\equiv {\rm erf}(x/\sqrt{2})$ and $E_2(x)\equiv \sqrt{2/\pi}\,e^{-x^2/2}$.  When $x\gg 1$ then $E_1(x)\to 1$, $E_2(x)\to 0$ and $1/x\ll 1$ so the $\mu_k$ become linear combinations of simple polynomials.

\subsection{Relation to simple polynomials}
In previous LP analyses, simple polynomials have been used to fit correlation functions.
For integer $n$, $L_n^{(\alpha)}$ is just a polynomial of order $n$, so one can also express $x^n$ as a linear combination of Laguerres:  
\begin{equation}
 \frac{x^n}{n!} = \sum_{j=0}^n (-1)^j {n+\alpha \choose n-j} \, L_j^{(\alpha)}(x).
 \end{equation}
Therefore, if one has fit $\xi_{\rm NL}$ to a simple polynomial, it is straightforward to transform those coefficients into those which would result from fitting to $n$th order Laguerre polynomials instead.  Hence, provided one accounts for the covariances between the fitted coefficients, the shape of the best fitting function will be the same.  In the main text we instead fit to $n$ {\em half-}integer Laguerre functions, because these are the functions which are singled out by Gaussian convolution, and for which the covariance matrix of the fitted coefficients can be easily used to provide error bands on the deconvolution/reconstruction.

\subsection{Centered Laguerre functions}\label{sec:centered}
The Laguerre reconstruction algorithm is designed to be used over the range of scales of order 100$h^{-1}$Mpc which are close to the BAO feature, the amplitude of which is small. However, for $x\gg 1$, the $\mu_k(x)$ can be large, and the best-fitting coefficients can have different signs, so the small amplitude of the correlation function at BAO scales is the result of large cancellations. 
Therefore, to avoid numerical inaccuracies, it is preferable to work with centered values.

We do so by subtracting a fiducial scale $r_{\rm fid}$ from all $r$ before fitting the model.  I.e., we replace Eq. (\ref{xiLpoly}) with
\begin{equation}
 \xi_{\rm L}(r) = \sum_{k=0}^n a_k\,\Bigl(\frac{r - r_{\rm fid}}{\sigma}\Bigr)^k .
 \label{xiLpoly-cen} 
\end{equation}
Integrating this over the Gaussian smearing kernel yields
\begin{equation}
  \xi_{\rm NL}(s) = \sum_{k=0}^n a_k\, \Big(\frac{\Sigma}{\sigma}\Big)^k\, \nu_k(x),
  \label{xiNLpoly-cen}
\end{equation}
where
\begin{equation}
 \nu_k(x) = \sum_{l=0}^k {k \choose l} \left(-\frac{r_{\rm fid}}{\Sigma}\right)^{k-l}\, \mu_l(x)
\end{equation}
and the $\mu_l(x)$ are the ordinary (non-centered) moments that appear in Eq.~(\ref{xiNLpoly}).

Since the $\nu_k$ are just linear combinations of the $\mu_k$, the result of fitting Eq.~(\ref{xiNLpoly-cen}) to the data must yield the same best-fit curve as when $r_{\rm fid}=0$.  In particular, this means that $r_{\rm LP-pre}$ and $r_{\rm LP-rec}$ should not -- and we have checked that they do not -- depend on the choice of $r_{\rm fid}$.  The only difference is that the coefficients of the fit are now better behaved, and the covariance matrix of the fitted coefficients is more stable.

\begin{figure}[b]
 \centering
 \includegraphics[width=0.9\hsize]{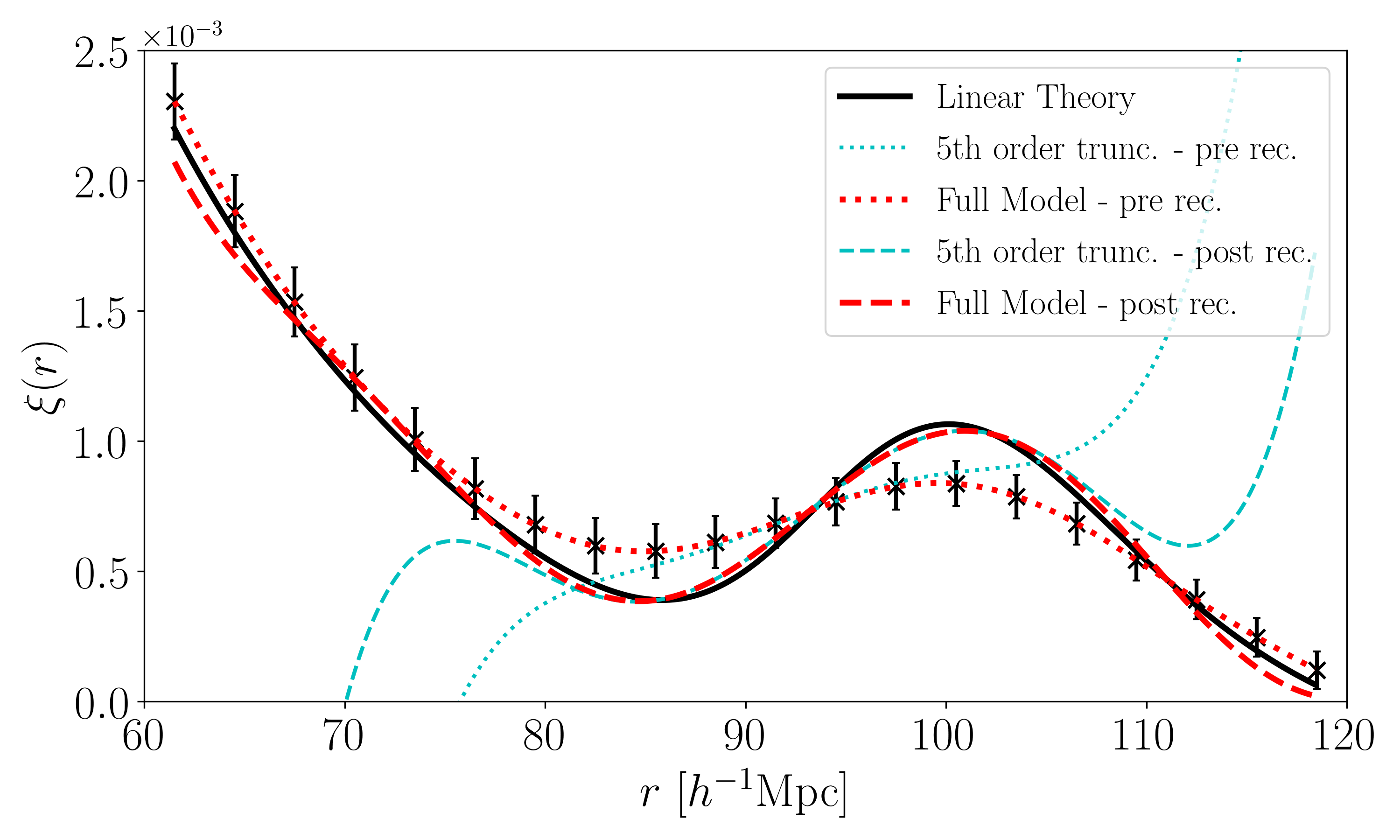}
 \caption{\label{fig:centered} Contribution of the first six terms to the correlation of the dark matter pre- and post-reconstruction (dotted and dashed cyan curves), when using centered functions (Eqs.~\ref{xiNLpoly-cen} and~\ref{xiLpoly-cen} with $r_{\rm fid}=r_{\rm LP - rec}$) and $\sigma=\Sigma = 4.6h^{-1}$Mpc).  Red curves show the sum of all ten terms.  Centering ensures that the lower order terms dominate on scales between the peak and dip; this is particularly evident post-reconstruction.}
\end{figure}

Therefore, in practice, having initially estimated $r_{\rm LP-rec}$ using some $r_{\rm fid}$, we set $r_{\rm fid} = r_{\rm LP-rec}$ and rerun the fitting routine.  While this again makes no difference to the shape of the resulting best fit curve, the coefficients of the associated reconstructed $\xi_{\rm Lag}$ are now more intuitive.  As $\xi_{\rm Lag}$ is now a simple polynomial centered on $r_{\rm LP-rec}$, only the lowest order terms contribute when $r-r_{\rm LP-rec}\ll\sigma$, as Fig. \ref{fig:centered} illustrates.  Symbols with error bars show the measured $\xi_{\rm NL}$, the dotted red curve shows the best fit to it with $n=9$ in Eq.~(\ref{xiNLpoly-cen}), the dashed red curve shows the reconstruction, $\xi_{\rm Lag}$ (Eq.~\ref{xiLpoly-cen}), and the solid black curve shows the linear theory $\xi_{\rm Lin}$.  The dotted and dashed blue curves show the result of truncating the sums in Eqs.~(\ref{xiNLpoly-cen}) and (\ref{xiLpoly-cen}) at $n=5$.  Evidently, the higher-order terms matter little between the peak and dip scales, suggesting that working with centered values is sensible.

\begin{figure}
 \centering
 \includegraphics[width=0.9\hsize]{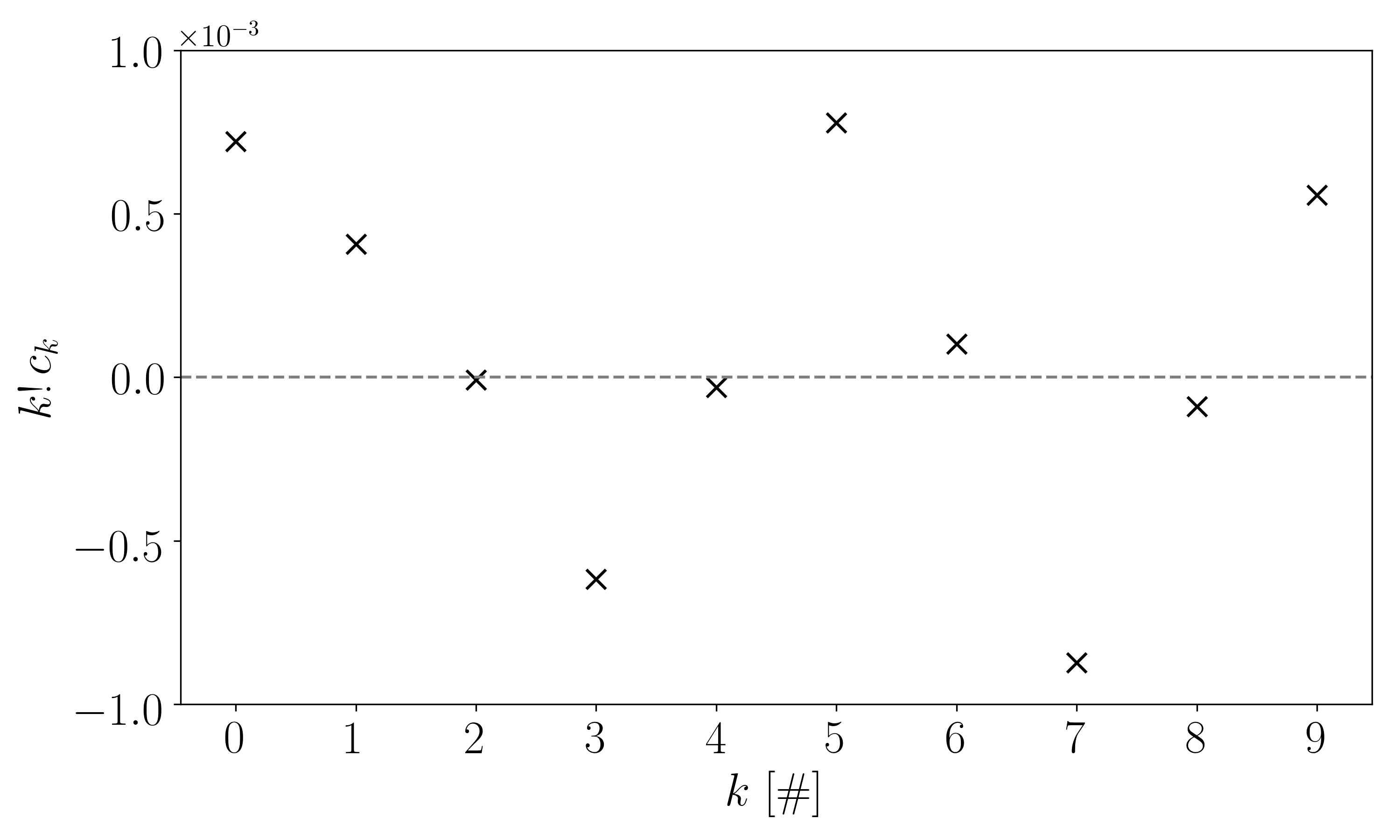}
 \caption{\label{fig:cent-coeff} 
Best-fit coefficients of the centered functions:  even coefficients are much smaller than odd ones, indicating that $\xi_{\rm Lag}$ is approximately an odd function around $r_{\rm LP-rec}$.
 }
\end{figure}

To make the point that the coefficients of the centered functions are intuitive, Fig.~\ref{fig:cent-coeff} shows $k!\,c_k$.  Except for $c_0$, which shifts the curve vertically without affecting its shape, the even coefficients are much closer to zero than the odd ones, indicating that $\xi_{\rm Lag}$ is approximately an odd function around $r_{\rm LP-rec}$.  The fact that scaling by $k!$ makes the odd coefficients approximately the same, but oscillating in sign, indicates that the odd function is approximately sinusoidal close to $r_{\rm LP-rec}$, as is readily apparent from looking at the shapes of $\xi_{\rm Lag}$ and $\xi_{\rm Lin}$ (by coincidence $(r_{\rm pk} - r_{\rm dip})/\pi \approx \Sigma$, so no further scaling was necessary to see this correspondence).

\section{Measurement details}\label{fitDetails}
As discussed extensively in \cite{PRDmocks}, we must make a number of choices when fitting a polynomial to the measurements:  these include the order of the polynomial to be fit, the range over which to fit, and the bin size (hence the number of bins to be fit).  We discuss the bin size first.

\begin{figure*}
 \centering
 \includegraphics[width=0.9\hsize]{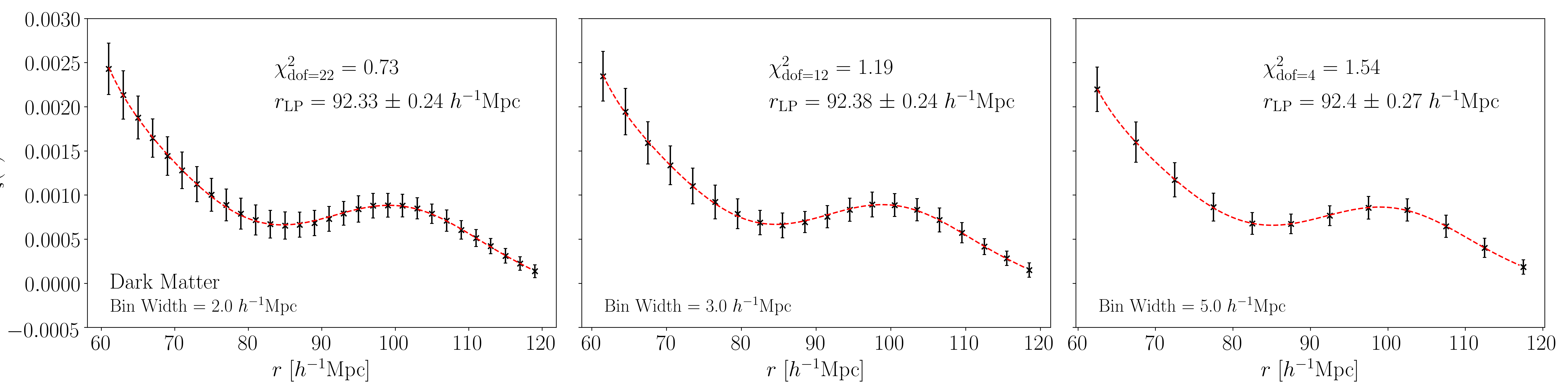}
 \includegraphics[width=0.9\hsize]{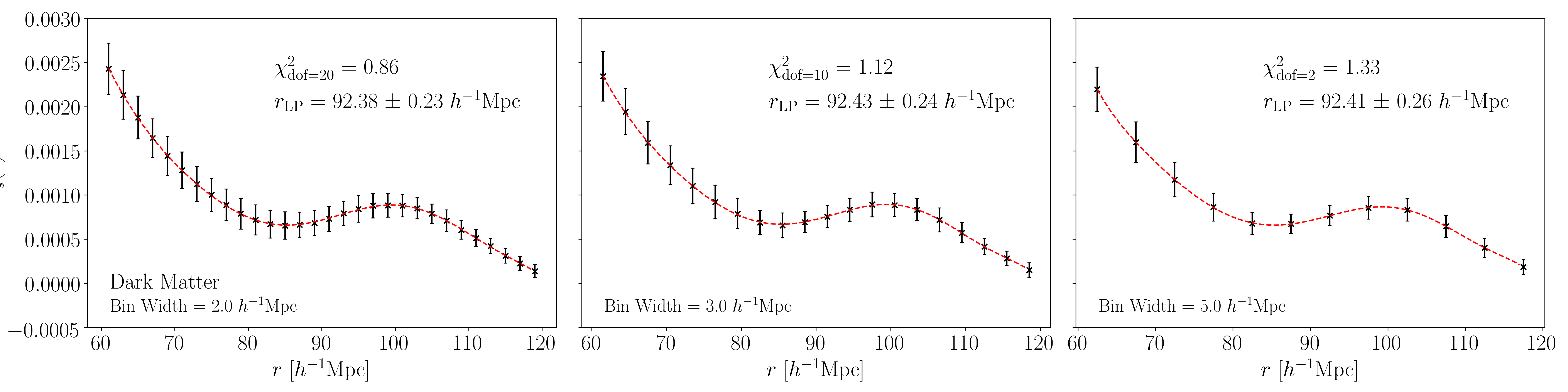}
 \caption{\label{fig:DM} 
   Dependence of goodness-of-fit and estimated $r_{\rm LP}$ on bin size (left to right) for a 7th-order (top) and 9th-order (bottom) $\mu_n$ function (dashed) fit to $\xi_{\rm NL}$ of the dark matter (symbols with error bars).  
 }  
\end{figure*}

\subsection{Dependence on bin width}
If the unbinned function is a polynomial, then correcting for the bin size is straightforward.  To see this, let $\xi_0(r)$ denote the correlation function in bins of vanishingly small size.  Then the correlation function in logarithmic bins of width $\epsilon$ is 
\begin{equation}
 \xi_\epsilon(r) = \frac{V_+\bar\xi(r_+) - V_-\bar\xi(r_-)}{V_+ - V_-},
\end{equation}
where  $V_\pm = (4\pi/3)\, r_\pm^3$,
       $r_\pm = r (1 \pm \epsilon/2)$
and 
\begin{equation}
 \bar\xi(r) = \frac{3}{r^3}\int_0^r {\rm d}x\,x^2\,\xi_0(x).
\end{equation}
If we parametrize $\xi_0$ using a polynomial,
\begin{equation}
 \xi_0(r) = \sum_{i=0}^n a_i\, r^i,
\end{equation}
then 
\begin{equation}
  \xi_\epsilon(r) = 
  \sum_{i=0}^n a_i \,\frac{3}{3+i}\,\frac{r_+^{3+i} - r_-^{3+i}}{r_+^3 - r_-^3}
  = \sum_{i=0}^n a_i\,r^i\,[1 + c_i(\epsilon)],
  \label{eq:xieps}
\end{equation}
since the term involving ratios of the $r_+$ and $r_-$ factorizes into the product of $r^i$ and a function of $\epsilon$.  From this it is obvious that extrema and inflection points of $\xi_\epsilon$ will not, in general, coincide with those of $\xi_0$.  The bias will depend on $\epsilon$, but also on the shape of $\xi_0$ (i.e. on the $a_i$).  (E.g., if $\xi_0$ has a feature -- a peak or dip -- that is narrower than $\epsilon$ then wide bins are more likely to lead to a bias.)

However, if we fit the measured correlation function to 
\begin{equation}
  \xi_\epsilon(r) = \sum_{i=0}^n b_i\, r^i,
\end{equation}
then the fitted coefficients $b_i$ are related to the intrinsic coefficients $a_i$ we want by 
\begin{equation}
  a_i = \frac{b_i}{1 + c_i(\epsilon)}.
\end{equation}
This shows that if $\xi_0$ is well described by a polynomial, then it is straightforward to correct for the bias induced by non-zero $\epsilon$ (i.e. logarithmic bins).  Keeping only the leading order terms in $\epsilon$ yields 
\begin{equation}
 c_i = \frac{\epsilon^2}{24}\, i(3+i);
\end{equation}
the scaling with $\epsilon^2$ rather than $\epsilon$ is why, in practice, the bin size effect is small.  For linear rather than logarithmic bins, Eq.(\ref{eq:xieps}) remains valid, but now $\epsilon = \Delta r/r$ for some constant $\Delta r$.  As a result, the $c_i$ depend on $r$.  While this makes it more complicated to reconstruct the $a_i$ from the $b_i$, correcting the bias is still possible.  

In practice, our bins are sufficiently small that these corrections are not necessary, but we have included this analyis to illustrate another useful property of a polynomial parametrization of $\xi$.  See \cite{smallr} for why polynomials are useful in the small-$r$ limit.

The discussion above shows that it would be useful to have an estimator of the correlation function which does not require binning.  Such estimators have recently become available \cite{lsqXi18, nobinXi21}.  These parametrize the correlation function in terms of basis functions.  Our work suggests that, in the BAO context, half-integer generalized Laguerre functions are a particularly useful choice.

\begin{figure*}
 \centering
 \includegraphics[width=0.9\hsize]{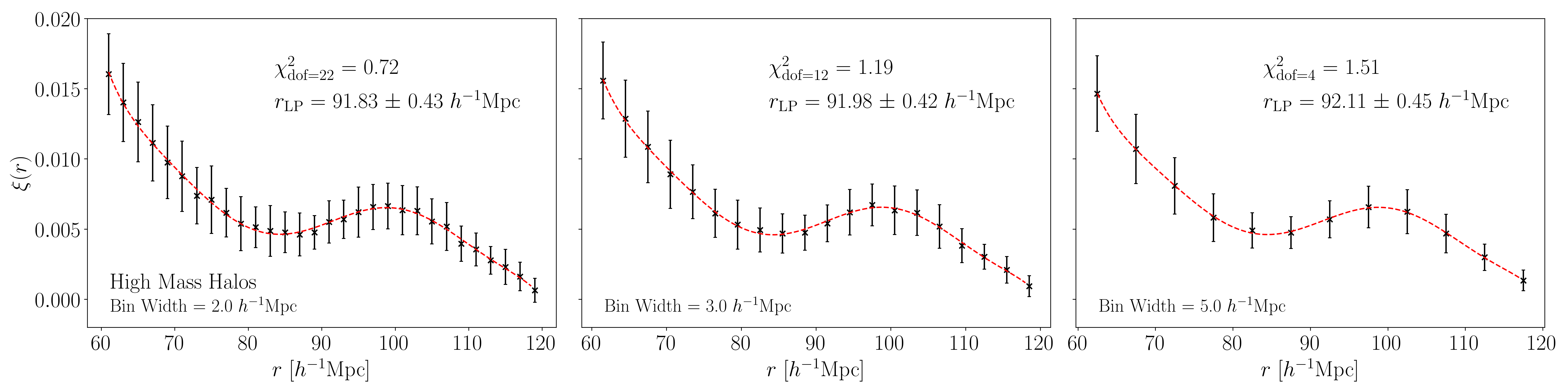}
 \includegraphics[width=0.9\hsize]{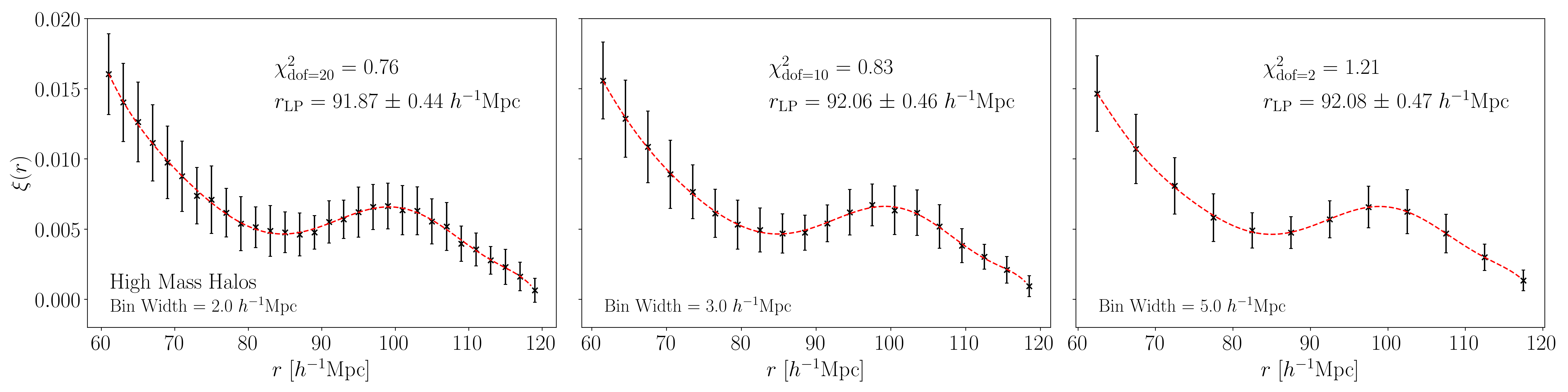}
 \caption{\label{fig:HM} 
   Same as previous figure but for high mass halos.}  
\end{figure*}

\subsection{Other sample-dependent choices}
Ref.~\cite{PRDmocks} shows that the optimal choices for estimating the BAO scale depend on the dataset (tracer number density and survey volume) but that, typically, one is only interested in the range that is within about 20$h^{-1}$Mpc on either side of the BAO feature, and the polynomial should have order $n\ge 5$.  E.g., in \cite{PRLboss} $n=5$ was sufficient, but in \cite{LPnus} $n=8$.


We have repeated the tests described in Ref.~\cite{PRDmocks} and verified that the same choices which apply when fitting an $n$th order polynomial also apply for the Laguerre functions which we describe and use in the main text.  These suggest that the range 75-115$h^{-1}$Mpc is nearly optimal.  However, because reconstruction is basically deconvolution, one wants the edges of the fitted region to be as far from the scales of interest as possible -- certainly  more than one smearing scale from the peak and dip scales.  We have found that fitting over the range 60-120$h^{-1}$Mpc produces no significant difference in the estimated $r_{\rm LP-pre}$, but returns significantly better reconstructions.  All the results in this paper use this 60-120$h^{-1}$Mpc range.

The fitting uses the full covariance matrix of the errors on the measurements.  As we note in the main text, we use an analytic estimate of this which includes both Poisson/discreteness and cosmic variance contributions.  The cosmic variance contribution requires a fiducial power spectrum and an estimate of the bias factor, but our results are not very sensitive to these choices.  E.g., there is no significant change to our results if we multiply the fiducial power spectrum by a smearing function $\exp(-k^2\sigma^2)$ or not, where $\sigma$ is the fiducial value described in the main text. (We have also compared, but do not show, results obtained using only the diagonal elements of this matrix with those which use the full matrix.)  Figures~\ref{fig:DM} and~\ref{fig:HM} show the results.  In each figure, comparison of the top and bottom panels shows that going to 9th-order in $\mu_n$ almost always returns $\chi^2/$d.o.f. closer to unity than just 7th order (we set the number of degrees of freedom equal to the number of bins minus the number of parameters to be fit), and that bins of width $3h^{-1}$Mpc are the most reliable.  

Therefore, in the main text we use the fits based on the full covariance matrix when fitting terms upto $\mu_9$ to measurements in bins of width $3h^{-1}$Mpc (i.e. the central panel in the bottom row of each figure).  Note, however, that the different choices explored in this Appendix only shift $r_{\rm LP}$ by less than the size of the quoted error bar.  Hence, the demonstration in the main text that $r_{\rm LP}$ shifts systematically with halo mass is robust against reasonable changes in the details of the fitting procedure.  

\section{The ABACUS+Emulator simulation set}\label{sec:AE}
The main text shows results that are based on an analysis of 20 realizations of the ABACUS simulation set.  However, the ABACUS suite includes 16 additional realizations of the same cosmological model that we will refer to as the Emulator set.  The only difference between the two sets is the choice of force-softening:  the original 20 simulations use Spline softening, whereas the Emulators use Plummer softening.  Ref.\cite{abacus} argue that, although spline softening is more accurate, the difference should be irrelevant for BAO studies.  Indeed, in their BAO work, \cite{baoHOD} use a combined Abacus + Emulator sample to arrive at an effective volume of 48~($h^{-1}$Gpc)$^3$.

\begin{table}
  \vspace{0.5cm}
  \begin{center}
  \begin{tabular}{c|c|c|c}
     \hline
    Tracer & $b_{10}$ & $r_{\rm LP-pre}$ & $r_{\rm LP-rec}$ \\ 
     \hline
      DM &  1  & $92.19\pm 0.12$ & $93.03\pm 0.11$ \\
      LM & 1.3 & $92.15\pm 0.13$ & $93.08\pm 0.13$ \\
      HM & 2.6 & $91.23\pm 0.26$ & $92.97\pm 0.24$ \\
     \hline
    \end{tabular}
\end{center}
\caption{Same as Table~\ref{tab:rLPs} in the main text, but now for $r_{\rm LP}$ only, in the combined Abacus+Emulator sample, an effective comoving volume of nearly 48~$h^{-3}$Gpc$^3$. }
  \label{tab:AE}
\end{table}

To enable a more direct comparison of our analysis with that in \cite{baoHOD}, we here perform all the analyses described in the main text on the combined Abacus and Emulator sample.  Table~\ref{tab:AE} shows the results. (The fits have similar $\chi^2$/d.o.f. to those in the main text.)  The most noteworthy difference with respect to the Abacus-only results in Table~\ref{tab:rLPs} is that the estimated $r_{\rm LP}$ scale in the combined Abacus+Emulator suite shows much larger shifts from linear theory and a stronger dependence on halo mass.

The final column in Table~\ref{tab:AE} shows that, despite the bigger shifts with respect to linear theory,  our reconstruction algorithm still works well.  In fact, comparison with the middle panel of Fig.2 in \cite{baoHOD} shows that our reconstructed precision of $\sim 0.15\%$ for the DM is comparable to that for the traditional, more elaborate, reconstruction schemes.

\begin{figure}
 \centering
 \includegraphics[width=0.9\hsize]{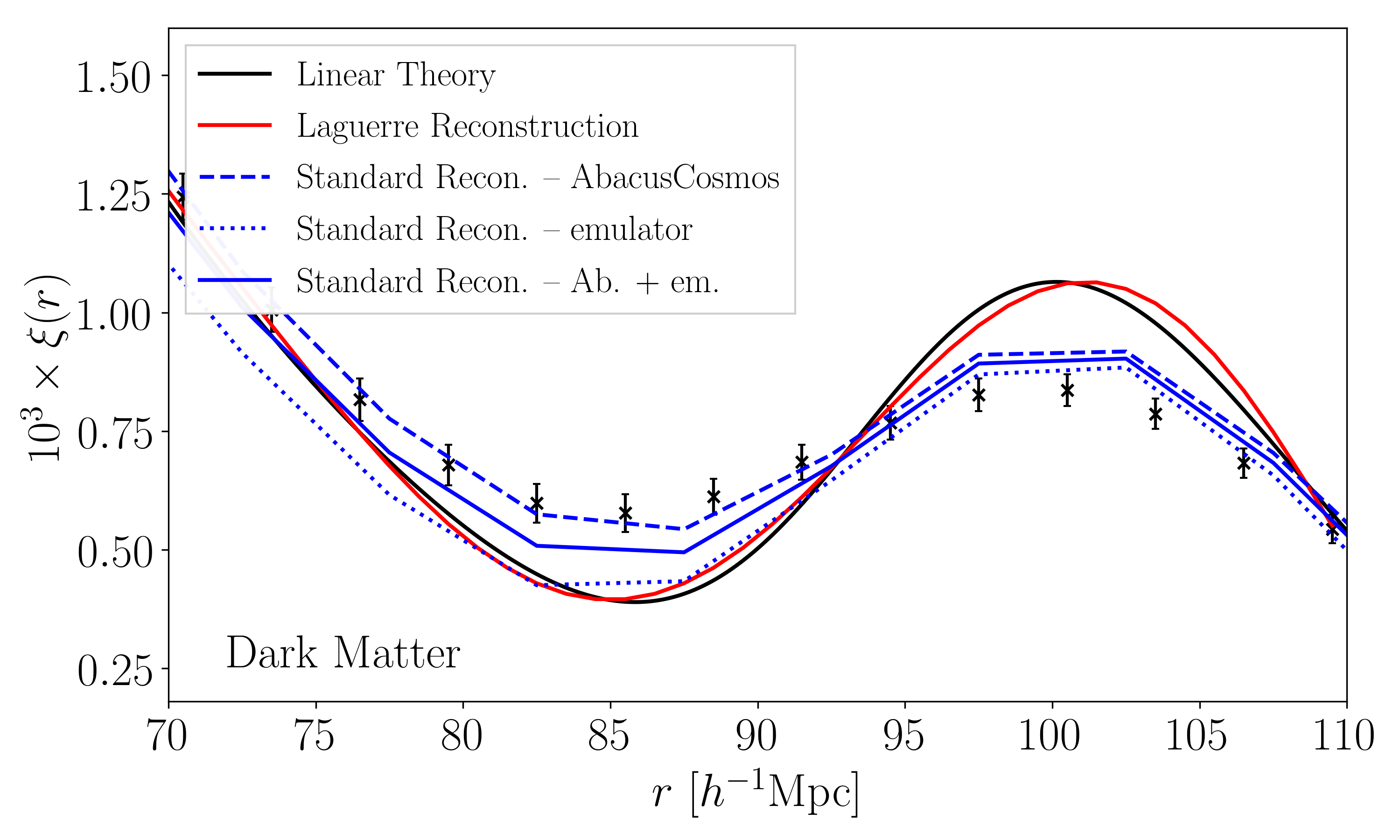}
 \caption{\label{fig:stdAE} 
   Same as Fig.~\ref{fig:std}, but for the combined Abacus + Emulator sample, and we only show the `standard' reconstruction after normalizing to match linear theory at 70$h^{-1}$Mpc.  Dashed and dotted curves show the contributions from the individual Abacus and Emulator simulation sets.}
\end{figure}

\begin{figure}
 \centering
 \includegraphics[width=0.95\hsize]{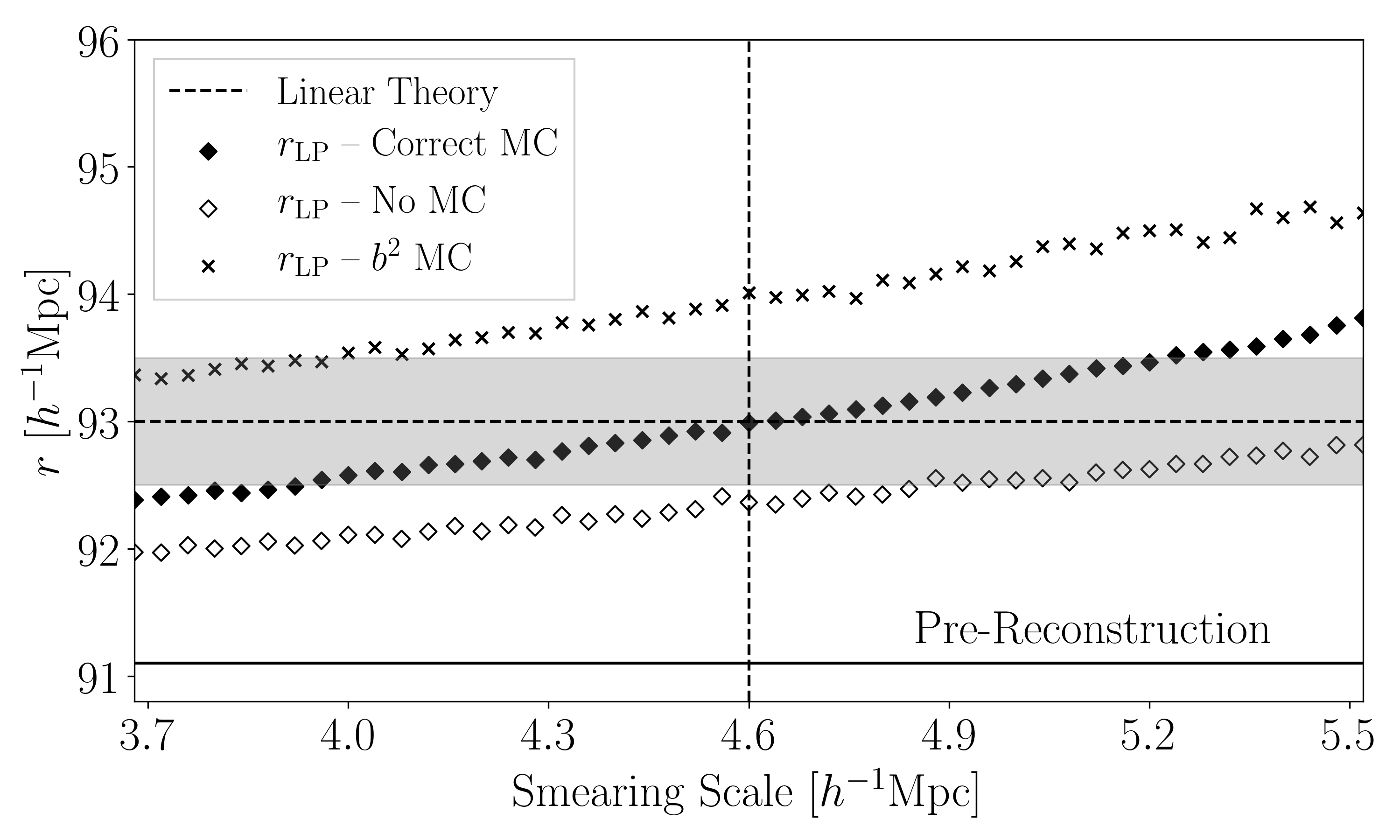}
 \caption{\label{fig:noMCAE} 
   Same as Fig~\ref{fig:noMC} but for the high mass halos in the combined
   Abacus+Emulator set.  Our reconstruction algorithm still works well,
   provided that we include the mode-coupling term (filled symbols).  
   Ignoring mode-coupling (open symbols) is substantially closer to
   linear theory, but not as close as in Fig.~\ref{fig:noMC}.
 }
\end{figure}

Fig.~\ref{fig:stdAE} -- similar to Fig.~\ref{fig:std} of the main text -- compares our Laguerre reconstruction with the `standard' reconstruction provided by \cite{baoHOD}.  The agreement with the linear theory shape is impressive.  While this is reassuring, our reconstruction works well  because the mode-coupling piece plays a significant role:  in Fig.~\ref{fig:noMCAE} open symbols, which assume no mode-coupling, are further from linear theory than the filled symbols.  This is a qualitative difference with respect to the results in the Abacus-only simulations (compare Fig.~\ref{fig:noMC}).  

\begin{figure}
 \centering
 \vspace{0.5cm}
 \includegraphics[width=0.9\hsize]{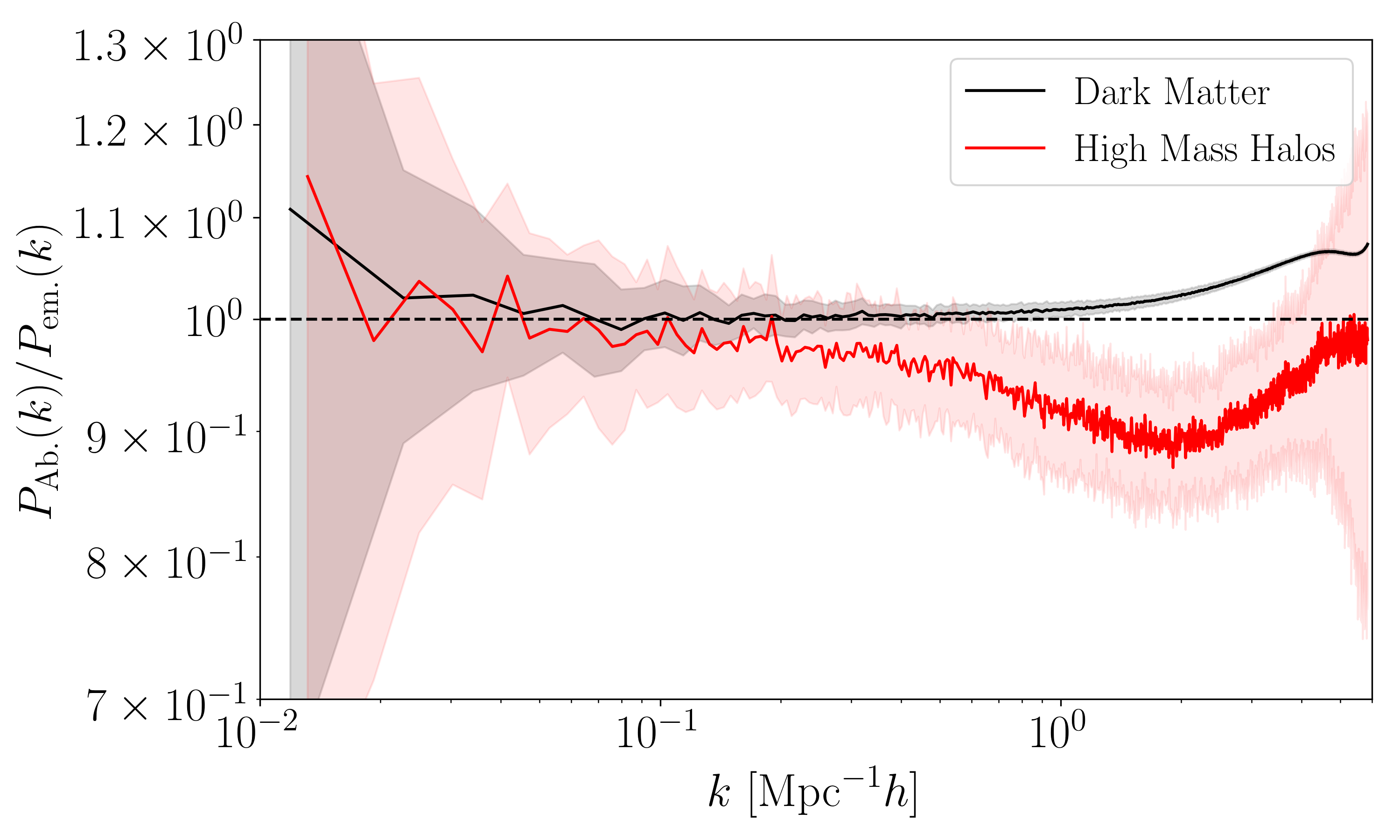}
 \caption{\label{fig:PkAE} 
   Comparison of evolved power spectra $P_{\rm NL}(k)$ in the Abacus and Emulator simulation sets shows good agreement for the dark matter, but can differ by up to ten percent for our massive halo (HM) sample.  Error bars show the measured rms scatter due to shot-noise and cosmic variance.  
 }  
\end{figure}

Presumably, these significant differences are due to differences in the shapes of $P(k)$ and $\xi(r)$.  (Indeed, the dashed and dotted curves in Fig.~\ref{fig:stdAE} show that the traditional `standard' reconstruction algorithm returns rather different shapes for the two sets.)  Fig.~\ref{fig:PkAE} shows that although $P(k)$ for the dark matter is in good agreement over scales relevant to BAO studies $k < 1h$/Mpc (consistent with Figs.4-7 in \cite{abacus}), the HM samples in the Emulator suite have slightly more power than their Abacus counterparts, especially at $k\gsim 0.3h$/Mpc.  The shaded bands show the scatter; the difference between the two simulation sets is difficult to explain with cosmic variance.

We also find that the comoving number density of the HM sample in the emulator set is about $0.96\times$ that in {\small ABACUS}, consistent with the small differences shown in Fig.2 of Ref.\cite{abacus}.  It is well known that there is a close connection between halo abundances and clustering \cite{st1999}.  Hence, because we define our samples using a fixed mass cut, we expect the Emulator sample to be slightly more strongly clustered.  Presumably this is what accounts for the small (few percent) approximately constant offset around $k~\sim 0.1h$/Mpc; differences in scale-dependent bias must contribute to the larger discrepancy at larger $k$.  The LM sample shows a similar level of discrepancy, both in terms of abundance and clustering strength.

\begin{figure}[b]
 \centering
 \includegraphics[width=0.9\hsize]{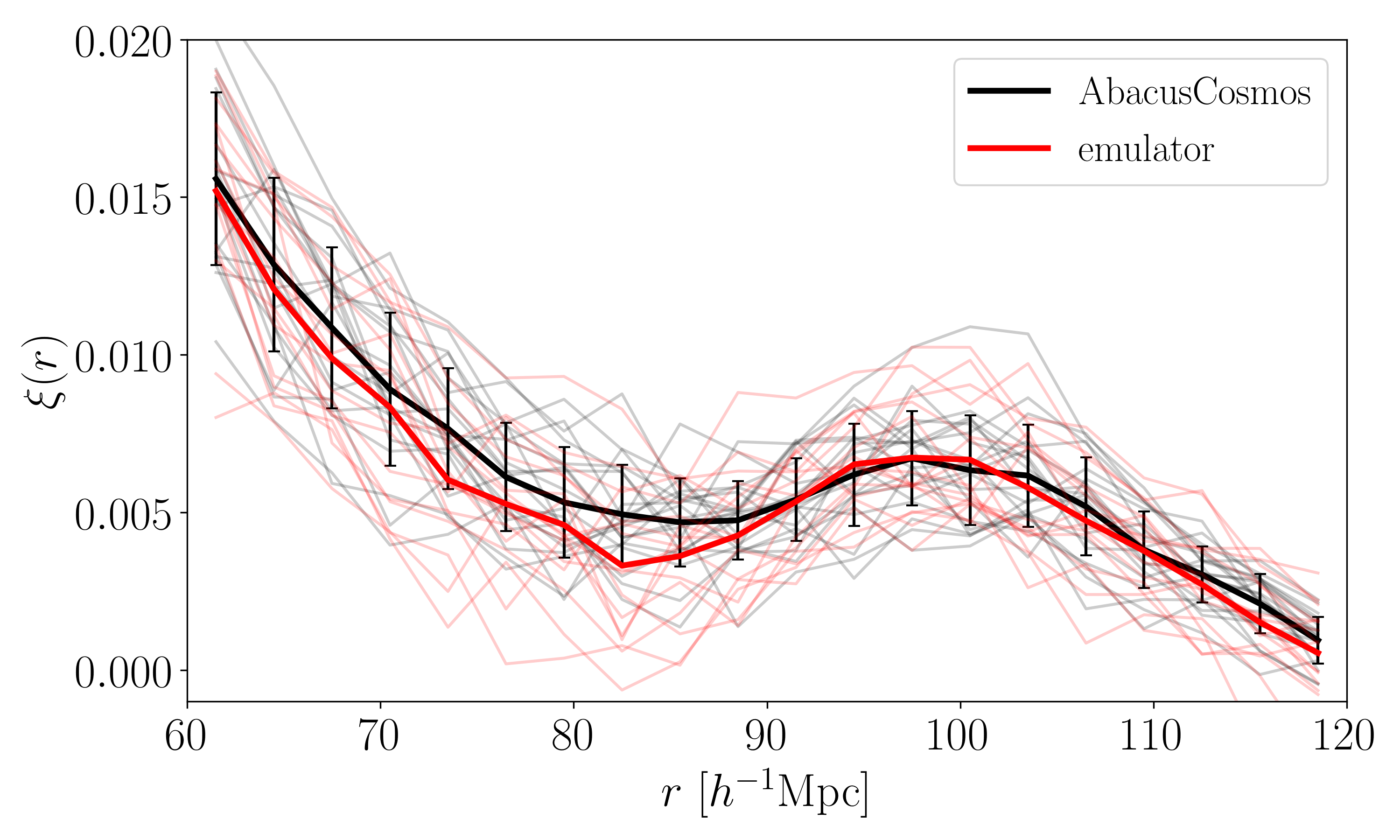}
 \caption{\label{fig:xiAE} 
  Evolved correlation functions $\xi_{\rm NL}(r)$ of the HM samples in the Abacus and Emulator simulation sets.  Thick curves show the ensemble-averaged value of each set. }  
\end{figure}

The question is:  Do these small differences matter?  Fig.~\ref{fig:xiAE} shows that the correlation functions of the two HM samples appear to have slightly different shapes, although the error bars (shown for Abacus-only) suggest that the difference may just be consistent with cosmic variance.  Since the $r_{\rm LP}$ methodology is supposed to be insensitive to shape differences arising from $k^2$-bias, it is possible that the Abacus and Emulator simulation sets each give consistent estimates of $r_{\rm LP}$, but combining their correlation functions leads to a bias (for the same reason that one can estimate the distance scale from blue and red galaxies separately, but one should {\em not} work with a curve that is the average of the two correlation functions).

With this in mind, we performed all the analyses described in the main text on the Emulator-only simulations.  The HM Emulator-only sample returns $r_{\rm LP} = 90.76\pm 0.46~h^{-1}$Mpc, compared to $92.06h^{-1}$Mpc for the Abacus-only sample in the main text.  In fact, a careful look at Fig.~\ref{fig:xiAE} shows that, even by eye, one would have guessed that the Emulator $r_{\rm LP}$ would be shifted to smaller scales (the peak and dip scales are both smaller).  The difference is substantially larger than the error bars, which we believe account for cosmic variance between the Abacus and Emulator suites.  Therefore, we do not understand the origin of these differences.  However, we {\em do} know that the Abacus spline-softening is more accurate \cite{abacus}.  This is why, in the main text, we only show results based on the more accurate Abacus simulations.

\end{document}